\documentclass[
  aip,
  author-year,
  author-year,
  reprint,
  onecolumn,
  tightenlines,
  12pt,
  amsmath,
  amssymb
]{revtex4-1}
\usepackage[pdftex]{graphicx}
\usepackage[pdftex,bookmarks,colorlinks,citecolor=blue,linkcolor=blue]{hyperref}
\usepackage[utf8]{inputenc}
\usepackage[T1]{fontenc}
\usepackage{siunitx}
\usepackage{bm}
\usepackage{url}
\setcitestyle{square}

\newcommand\secref[1]{Sec.~\ref{#1}}
\newcommand\appref[1]{Appendix~\ref{#1}}
\newcommand\figref[1]{Fig.~\ref{#1}}
\newcommand{\id}{\bm{I}}
\newcommand{\phiDST}{\phi_{\mathrm{c}}}
\newcommand{\phiFricJam}{\phi_{\mathrm{J}}^{\mu}}
\newcommand{\phiJam}{\phi_{\mathrm{J}}^{0}}
\newcommand{\tanVec}{\bm{t}_{ij}}
\newcommand{\operatorPn}{\operatorname{P}_{\bm{n}_{ij}}}
\newcommand{\operatorPt}{\operatorname{P}_{\bm{n}_{ij}}'}
\newcommand{\operatorPr}{\operatorname{P}_{\bm{n}_{ij}}^{\mathrm{r}}}
\newcommand\ReNum{\mbox{\textit{Re}}}  
\newcommand\PeNum{\mbox{\textit{Pe}}}  
\newcommand\StNum{\mbox{\textit{St}}}  

\begin{document}

\title{Shear thickening, frictionless and frictional rheologies in non-Brownian suspensions}

\author{Romain Mari}
\affiliation{Benjamin Levich Institute, %
City College of New York, New York, NY 10031, USA}
\author{Ryohei Seto}
\affiliation{Benjamin Levich Institute, %
City College of New York, New York, NY 10031, USA}
\author{Jeffrey F. Morris}
\affiliation{Benjamin Levich Institute, %
City College of New York, New York, NY 10031, USA}
\affiliation{Department of Chemical Engineering, %
City College of New York, New York, NY 10031, USA}
\author{Morton M. Denn}
\affiliation{Benjamin Levich Institute, %
City College of New York, New York, NY 10031, USA}
\affiliation{Department of Chemical Engineering, %
City College of New York, New York, NY 10031, USA}

\date{\today}


\begin{abstract} 
Particles suspended in a Newtonian fluid raise the viscosity
and also generally give rise to a shear-rate dependent rheology.
In particular, 
pronounced shear thickening may be observed at large solid volume fractions.
In a recent article
(R. Seto, R. Mari, J. F. Morris, and M. M. Denn., Phys. Rev. Lett., 111:218301, 2013)
we have considered the minimum set of components 
to reproduce the experimentally observed shear thickening behavior, 
including Discontinuous Shear Thickening (DST).
We have found frictional contact forces to be essential, 
and were able to reproduce the experimental behavior by a simulation including this
physical ingredient along with viscous lubrication.
In the present article, 
we thoroughly investigate the effect of friction 
and express it in the framework of the jamming transition.
The viscosity divergence at the jamming transition 
has been a well known phenomenon in suspension rheology, 
as reflected in many empirical laws for the viscosity.
Friction can affect this divergence, and in particular the jamming packing fraction 
is reduced if particles are frictional. 
Within the physical description proposed here, 
shear thickening is a direct consequence of this effect: 
as the shear rate increases, 
friction is increasingly incorporated as more contacts form, 
leading to a transition from a mostly frictionless to a mostly frictional rheology.
This result is significant because it shifts the emphasis 
from lubrication hydrodynamics and detailed microscopic interactions 
to geometry and steric constraints close to the jamming transition. 
 \end{abstract}

\maketitle 

\section{Introduction}

\subsection{Shear thickening}
\label{sec:Intro_shearthickening}

Suspensions (solid particles immersed in a fluid) 
exhibit a wide range of rheological behaviors,
including shear thinning, shear thickening,
and finite normal stress differences.
Shear thickening~\citep{Barnes_1989, Mewis_2011, Brown_2014}, 
where the viscosity increases with shear rate
even if the suspending fluid is Newtonian,
is a particularly intriguing phenomenon.
In the extreme case of Discontinuous Shear Thickening
(DST; for early descriptions see the work of
\citet{Williamson_1930, Williamson_1931, Freundlich_1938}), 
which is observed at high volume fractions of solid material, 
the viscosity can increase by
several orders of magnitude at a critical shear rate. 
Note that we do not include here the irreversible shear thickening
occurring due to particle aggregation (as seen for example in fumed
silica~\citep{Crawford_2013}) under the name DST.


The variety of systems showing a DST suggests that 
this may be a universal behavior
that is obtained with a minimum set of physical ingredients.
Even though most of the data available are for Brownian suspensions
(that is, sub-micrometer particles)~%
\citep{
  Metzner_1958, Hoffman_1972, Bender_1995, Bender_1996, Frith_1996,
  Fagan_1997, Boersma_1990, DHaene_1993, OBrien_2000, 
  Maranzano_2001, Maranzano_2001a, Maranzano_2002},
thermal motion does not seem necessary to observe DST, 
as experiments with micrometer scale~%
\citep{Boersma_1990, Lootens_2004, Lootens_2005, Larsen_2010} 
or larger particles~%
\citep{Freundlich_1938, Bertrand_2002, Brown_2009, Brown_2012,
Fall_2010, Fall_2012} show.
Inertia has also been associated with the existence of shear thickening in suspensions~%
\citep{Lemaitre_2009, Fall_2010, Trulsson_2012, Fernandez_2013, Kawasaki_2014}, 
based on the initial ideas of \citet{Bagnold_1954},
but the Bagnoldian scaling of a viscosity proportional to the shear rate is
clearly milder than the abrupt shear thickening observed in experiments. 
More importantly the thickening due to inertia arises at Stokes number
$\StNum \equiv \rho a^2 \dot{\gamma}/\eta_0$
(with $\dot \gamma$ the shear rate, $\eta_0$ the viscosity of the fluid phase,
and $\rho$ and $a$ the mass density and size of the solid particles, respectively)
of order $\StNum \approx 1$ in the simulations~%
\citep{Trulsson_2012, Fernandez_2013, Kawasaki_2014},
which contrasts with the values of at most $\StNum \approx 10^{-3}$
found in experiments (see for example~\citep{Maranzano_2001, Fall_2010, Brown_2012}).
Thus inertia is not necessary for DST in the conditions
probed by rheometric flows.
(Flows involving significantly larger particles,
say in the \si{\milli\meter} range,
or significantly larger shear rates $\dot \gamma \gg \SI{1}{\per\second}$,
can probe the regime where inertia matters).
%

Indeed, the apparently simple experimental system
of nearly rigid non-Brownian neutrally buoyant particles 
immersed in a Newtonian fluid exhibits DST in the Stokes regime~%
\citep{Brown_2009,Brown_2012},
indicating that the phenomenon stems from a rather restricted
set of simple ingredients.
The puzzle has very few pieces.
In the Stokes regime, however, non-Brownian neutrally buoyant hard
spheres in a Newtonian fluid will create a suspension whose rheology
is independent of shear rate, as there is only one force scale,
the hydrodynamic one (we will further
explain this fact in~\secref{sec:shearrate_dep}).
So there cannot be \emph{too few} pieces.

\subsection{Fluid mechanics and granular physics perspectives}

The flow of suspensions has historically been studied from a fluid mechanics perspective, 
and the emphasis has been on a description based on hydrodynamic interactions.
Suspensions are usually described as hard particles immersed in a Newtonian fluid,
interacting through hydrodynamics (including Brownian forces) 
and sometimes an additional soft repulsive potential 
(approximating an electric double-layer or mimicking polymer coating, for instance).
They are typically studied in the Stokes flow regime, 
at vanishing particle Reynolds number 
$ \ReNum \equiv \rho_0 a^2 \dot{\gamma}/\eta_0$
(with $\rho_0$ the mass density of the fluid phase) and Stokes number.
%
The key point is that the hard cores of the particles
are treated as boundary conditions for the Stokes equations
that describe the fluid phase;
the particles never directly generate forces through contacts.
This treatment is self-consistently justified by the fact that the Stokes flow 
between two rigid surfaces leads to a lubrication force
whose resistance coefficient diverges at contact,
effectively preventing two particles from colliding.
Within this framework, shear thickening is explained by the creation
at large shear rates of locally denser clusters of particles (\emph{hydro-clusters}),
which are highly dissipative due to the singular
lubrication flows between the particles~%
\citep{Brady_1985, Bender_1996, Melrose_2004, Wagner_2009}.
While this purely fluid mechanical point of view is able to describe
the flow and rheology of moderately concentrated suspensions,
it seems unable to explain the abrupt shear thickening observed in highly concentrated suspensions.
Simulations by Stokesian Dynamics 
give a weak logarithmic shear thickening~%
\citep{Brady_1988, Bossis_1989, Phung_1996, Foss_2000, Melrose_2004a}, 
thereby raising the issue of the validity 
of this approach at high volume fractions.
Moreover, DST is also observed in athermal suspensions, where the
shear rate dependence introduced by the Brownian force scale
disappears: in this regime, a purely hydrodynamic perspective would
predict a shear-rate independent rheology.


In the past few years,
new ideas have emerged from the granular rheology perspective.
\citet{Boyer_2011} developed an analogy between the rheology of suspensions 
and the rheology of granular materials.
In dry granular flow, the rheology depends on 
the ratio between inertia and particle pressure~\citep{Deboeuf_2009}:
if the inertia dominates, particles bounce around and the stresses are
essentially due to momentum exchange based on collisions
(a ``granular gas'' regime), whereas if the pressure dominates, particles are forced
to stay at contact and the stresses are dominated by contact force chains
(a ``granular packing'' regime).
This perspective can be incorporated in 
a dimensionless \emph{inertial number}
$I \equiv \dot{\gamma} d \sqrt{\rho/\Pi}$~\citep{da-Cruz_2005},
where $d$ is the diameter of the solid particles, 
$\rho$ their density, and $\Pi$ the particle pressure.
Similarly, for suspensions in the Stokes regime
there is a \emph{viscous number} $I_{\mathrm{v}} \equiv \eta_0 \dot\gamma/\Pi$, 
where $\eta_0$ is the viscosity of the suspending fluid,
which compares viscous dissipation to the particle pressure~\citep{Boyer_2011}.
The stresses can be dominated either by viscous dissipation 
if $I_{\mathrm{v}} \gg 1$ 
(when particles are far apart) 
or by a contact network if $I_{\mathrm{v}} \ll 1$.
This results in constitutive laws for the viscosity 
and other stress components
and for the volume fraction 
that are unique functions of $I_{\mathrm{v}}$.
One can anticipate that the regimes $I \ll 1$ and $I_{\mathrm{v}} \ll
1$, both of which are dominated by the approach to the jamming transition,
will share some similarities.
Indeed, they do share a power-law scaling (typical of the jamming transition) 
between particle pressure and distance to the jamming point 
$\Pi \sim (\phi_{\mathrm{J}}-\phi)^{-\lambda}$~\citep{Forterre_2008, Boyer_2011}.
The scaling is the same for the other stress components,
including the important case of shear stress 
$\sigma \sim (\phi_{\mathrm{J}}-\phi)^{-\lambda}$. 
These power laws are now argued to be a generic behavior close to
jamming~\citep{Lerner_2012,Andreotti_2012},
though the value of the exponent $\lambda$ might be system dependent
(it might depend on shape or friction for instance).
With these ideas, the emphasis has shifted 
from hydrodynamics and detailed microscopic interactions 
to geometry and steric constraints close to the jamming transition.


While those new ideas have proven to be successful in explaining
some of the rheology at high volume fractions~%
\citep{Boyer_2011, Trulsson_2012, Lerner_2012}, 
they cannot account for non-Newtonian behavior in the Stokes regime.
The case of DST is particularly puzzling: 
whereas intuition suggests that
DST is a manifestation of jamming~%
\citep{Cates_1998a,Bertrand_2002,Lootens_2003,Hebraud_2005,Brown_2009}, 
DST is not captured by the $I_{\mathrm{v}}$-based rheology,
which predicts no shear rate dependence in the Stokes regime~%
\citep{Lerner_2012, Trulsson_2012},
and reduces to a $\phi$-dependent rheology~\citep{Boyer_2011}.
Simply stated,
this rheology predicts the correct volume fraction dependence of the viscosity 
at fixed shear rate, 
but completely fails at predicting the shear rate dependence at fixed volume fraction.
Again, it is worth noticing that this absence of shear rate dependence
hints at a missing force (or time) scale in the description of dense
suspensions.

\subsection{This work}

In this article, 
which extends and complements a previous publication~\citep{Seto_2013a}, 
we show how jamming is related to shear thickening
and how to escape the apparent contradiction described above. 
The starting point is to note that the jamming transition,
while it exhibits some generic features 
that are independent of the microscopic interactions 
like the scalings described above, 
retains a clear signature of the microscopic details in the volume fraction at which it occurs. 
For instance, it is well known that jamming can occur anywhere between 
$\phi_{\mathrm{J}}^{\mu=\infty} \approx 0.55$ and
$\phi_{\mathrm{J}}^{\mu=0} \approx 0.64$ depending on the friction
coefficient $\mu$ between the grains~\citep{Silbert_2002,Jerkins_2008,Song_2008,Silbert_2010}.
This means that the relation $\phi(I_{\mathrm{v}})$ actually depends on $\mu$, 
which in turn implies that the viscosity $\eta(\phi)$ depends on $\mu$. 
Therefore, if there exists a mechanism such that $\mu$ (or an effective friction coefficient) 
varies with the shear rate $\dot{\gamma}$, 
one can obtain a non-Newtonian rheology within the framework of \citet{Boyer_2011}.
This mechanism has to be associated with an extra force scale besides
the hydrodynamic one in order to provide the shear-rate dependence.


We introduced such a mechanism in the recent 
article~\citep{Seto_2013a}, 
and here we thoroughly explore its consequences through numerical simulations of
dense (i.e., highly concentrated) non-Brownian suspensions under simple shear flow.
We note that
two other recent papers have independently introduced 
similar but somewhat different mechanisms 
to achieve a rate-dependent friction coefficient $\mu(\dot{\gamma})$
in simple shear flow simulations~\citep{Fernandez_2013, Heussinger_2013}.
Both of them prove that interparticle friction is essential to shear
thickening of dense suspensions.
In contrast to those works, the present work however show that inertia is inessential to this
phenomenon, as we obtain it in an overdamped simulation (see next
Section).
This is important because results coming from simulations including inertia
gives shear thickening at shear rates that are several orders of
magnitude larger than the experimental values (see previous~\secref{sec:Intro_shearthickening}).
We show here that our inertialess model gives a \emph{quantitative}
description of shear thickening (see Conclusion~\secref{sec:conclusion}).


The present work shows the importance of frictional contacts 
in dense suspension rheology.
Of course, 
particles must form contacts during flow
in order for the rheology to depend on the friction coefficient, 
and here we come up against one consequence of the fluid mechanics perspective.
In fact, it is known that particles do come in contact under flow, 
even in the dilute limit, 
perhaps due to surface roughness~%
\citep{Davis_1992,Zhao_2002,Lootens_2005,Blanc_2011a}.
One simply has to accept that those contacts, 
though not essential to the rheology of dilute or semi-dilute suspensions 
(as the successes of Stokesian Dynamics demonstrate),
become important at large volume fractions.
But contacts are a necessity if one thinks of the large volume-fraction limit of jamming 
($I_{\mathrm{v}} \to 0$): 
there is no more room in this limit
for lubrication films between particles, 
and contacts must proliferate and dominate the rheology. 
Besides enforcing the geometric constraint of no overlapping particles, those
contacts bring a new restriction to reorganization at the microscopic level:
they carry significant tangential forces due to friction.
It is worth noting that even though the close-range hydrodynamics
(i.e., lubrication forces) also generate some tangential forces, 
they are of a fundamentally different nature.
For a relative motion between two nearby particles, the normal lubrication
force diverges as the inverse of the interparticle gap, 
whereas the tangential lubrication force diverges only logarithmically with vanishing gap, 
which means that the effective friction coefficient vanishes in the limit of a small gap.
Thus lubrication, as in its literal meaning, 
provides very little resistance to relative tangential motion 
compared to relative normal motion.
This stands in contrast to the frictional contact forces, 
for which the friction coefficient is finite;
i.e., tangential and normal contact forces are of comparable scale.
It follows that the constraint introduced by contact friction is
much more efficient at increasing the viscosity than is lubrication.


Describing the physics of contact and motion
when the gap between two particles in a suspension is smaller than, 
say, \SI{1}{\nano\meter}, is not an easy task.
Moreover, one cannot expect to find generic behavior, 
as interactions will vary depending on the nature of the solid particles 
and of the suspending fluid.
However, this level of detail may not be essential 
to understanding
the physics that emerges at a macroscopic level. 
At high volume fractions
it is reasonable to expect that
the singularity associated with the jamming transition
will dominate the rheology of suspensions.
Our approach in this work is thus 
to study the rheology of minimal model systems
that include hydrodynamics, a repulsive interaction, 
and frictional contacts between particles. 
These interactions, while being realistic, will be simplified to their essence.

\section{Models and methods}

In this section, we will describe our numerical model of
a dense non-Brownian suspension under shear flow~\citep{Denn_2014}.
We intend to have a minimal model
containing as few physical elements as possible while exhibiting a DST.
Even if the model is intended to be as generic as possible,
a few parameter choices are necessary.
For those choices, we try to set values that correspond to a realistic
interpretation of a suspension of hard (say, silica) spheres with radius
in the \SIrange{1}{10}{\micro\meter} range,
density matched (or such that the sedimentation occurs over times much larger
than the inverse shear rate) and charge stabilized.
The only parameter that takes an unrealistic value is the particles' stiffness,
which is smaller in the simulations than in a typical experimental system.
A stiffness comparable to experimental values for silica (or even PMMA),
would require the use of very small time steps in the simulation
in order to resolve the dynamics of the contacts which correspond
to an overlap in the \si{\nano\meter} range. 
We however circumvent this problem by tuning the particle stiffness
with applied shear rate such that we get rid of particle
deformation effects in the rheology
(see \secref{sec:contact_model} and \appref{app:contact_model_full} for details).

\subsection{Merging hydrodynamic interactions and granular contact models}
\subsubsection{Equations of motion}

In a suspension, the flow of the fluid is described by the
Navier-Stokes equation, while the motion of
the particles is given by Newton's equation of motion:
\begin{equation}
  \bm{m}\cdot \frac{\mathrm{d}}{\mathrm{d} t} 
  \begin{pmatrix} 
    \bm{U}\\
    \bm{\Omega} 
  \end{pmatrix}
  = \sum_{\alpha} 
  \begin{pmatrix} 
    \bm{F}_{\alpha}(\bm{U}, \bm{\Omega}, \bm{r})\\
    \bm{T}_{\alpha}(\bm{U}, \bm{\Omega}, \bm{r})
  \end{pmatrix},
  \label{eq:eq_motion}
\end{equation}
where $\bm{m}$ is the mass/moment-of-inertia matrix, $\bm{U}$ and
$\bm{\Omega}$ are the translational and rotational velocity vectors, and
$\bm{F}_{\alpha}$ and $\bm{T}_{\alpha}$ are the force and torque vectors,
respectively.
The right-hand side consists of two different types of forces: 
some depend only on the configurations $\bm{r}$ of the particles 
(e.g., forces derived from a potential),
but some also depend on the velocities $\bm{U}$ and $\bm{\Omega}$,
including inelastic contact forces (e.g., contacts with dashpot terms)
and fluid-particle interactions (hydrodynamic interactions).


We will study the rheology of this suspension under an imposed simple
shear flow field $\bm{U}^{\infty}(\bm{r})$ expressed using the
vorticity $\bm{\Omega}^{\infty}$ and the rate-of-strain tensor $\bm{E}^{\infty}$ as
\begin{equation}
  \bm{U}^{\infty}(\bm{r})
  =  \bm{\Omega}^{\infty} \times \bm{r}
  + \bm{E}^{\infty} \cdot \bm{r}.
\end{equation}
At a shear rate $\dot{\gamma}$, a simple shear flow corresponds to the
following nonzero elements: $\Omega^{\infty}_3 = \dot{\gamma}/2$ and
$E^{\infty}_{12} = E^{\infty}_{21}= \dot{\gamma}/2$.


In many experimental flows, including the ones for which shear
thickening is typically observed, 
the particle Reynolds number and the Stokes number are very small
due to the small size of the suspended particles. 
This means that the equation of motion for the particles can be studied in
its overdamped limit, which is simply a quasi-static force balance equation.
As will be discussed
in Secs.~\ref{sec:hydrodynamic_intereactions},
\ref{sec:contact_model}, and \ref{sec:repulsion} below, 
in this regime the velocity dependence 
coming from the hydrodynamic interactions and the contact forces 
has a linear form, and the force balance
equation can be written as a linear algebraic relation having the form
\begin{equation}
 \bm{0} = \bm{R} \cdot
  \begin{pmatrix}
    \bm{U}-\bm{U}^{\infty}\\
    \bm{\Omega}- \bm{\Omega}^{\infty}
  \end{pmatrix}
+
 \begin{pmatrix}
    \bm{F}(\bm{r})\\
    \bm{T}(\bm{r})
  \end{pmatrix}
,
  \label{eq:fb}
\end{equation}
%

Therefore, our simulation of a suspension under a simple shear flow 
in the Stokes regime requires solving~\eqref{eq:fb} for the velocities,
and obtaining the positions $\bm{r}$ of the particles at any time $t$
through time integration of these velocities.

\subsubsection{Hydrodynamic interactions}
\label{sec:hydrodynamic_intereactions}

In principle, hydrodynamic interactions
can be determined by solving the Stokes equations,
but this is extremely expensive from a computational view point.
For dense suspensions, however, the dominant hydrodynamic interactions
come from the fluid flow in the narrow gaps between nearby solid
particles~\citep{Ball_1997}, because the resistance to relative motion
becomes singular when the gaps are
vanishingly small.
They give rise to pair-wise short range lubrication forces
(as opposed to the many-body nature of the full, long-range hydrodynamic interactions).
As a consequence, the linear relations between velocities and
hydrodynamic forces simply contain a contribution from the Stokes
drag and a contribution from the lubrication:
\begin{equation}
  \begin{pmatrix}
    \bm{F}_{\mathrm{H}} \\
    \bm{T}_{\mathrm{H}}
  \end{pmatrix}
  =
  - \bigl(\bm{R}_{\mathrm{Stokes}}+\bm{R}_{\mathrm{Lub}}\bigr)
  \cdot
  \begin{pmatrix}
    \bm{U}-\bm{U}^{\infty}\\
    \bm{\Omega}-\bm{\Omega}^{\infty}\\
  \end{pmatrix}
  + 
  \bm{R}'_{\mathrm{Lub}}:
  \bm{E}^{\infty}.
\end{equation}
$\bm{R}_{\mathrm{Stokes}} $ is a diagonal matrix
giving Stokes drag forces and torques, and $\bm{R}_{\mathrm{Lub}}$
and $\bm{R}'_{\mathrm{Lub}}$ are sparse matrices~\citep{Ball_1997}.


Consistent with our choice of keeping only physically
relevant near-distance interactions,
we use only the leading terms for the resistance matrices.
Physically, the terms included in our model correspond to
the squeeze, shear, and pump modes of \citet{Ball_1997}
(we do not consider their twist mode,
as it is not associated with a divergence of the resistance at contact,
and is thus subdominant).
Defining the non-dimensional gap $h^{(i,j)}$ 
between particles $i$ and $j$ having radii $a_i$ and $a_j$ 
as $h^{(i,j)} \equiv 2(r^{(i,j)} - a_i - a_j)/(a_i+a_j)$ 
and the center-to-center unit vector 
$\bm{n}_{ij} \equiv (\bm{r}^{(j)}-\bm{r}^{(i)})/r^{(i,j)}$ 
with $r^{(i,j)} \equiv |\bm{r}^{(j)}-\bm{r}^{(i)}|$, 
the modes associated with relative displacements respectively along 
and tangential to $\bm{n}_{ij}$ have singular behaviors with leading terms 
diverging as $1/h^{(i,j)}$ and $\log(1/h^{(i,j)})$~\citep{Jeffrey_1984,Jeffrey_1992}.
The consequence of the singularity in the squeeze mode (i.e., along $\bm{n}_{ij}$) 
is that there should not be any contact between particles~\citep{Ball_1995}.
However, as mentioned in the introduction, 
this statement is only true for perfectly idealized situations, 
and not realistic even for model experimental systems consisting 
of spherical particles due to, for example, a finite surface roughness~%
\citep{Davis_1992,Zhao_2002,Lootens_2005,Blanc_2011a}
or a breakdown of the continuity assumption for the fluid at the molecular mean
free path scale~\citep{Ho_1998}.
In order to mimic this reality, 
we regularize the singularities arising in the lubrication resistances 
by inserting a small cutoff length scale $\delta$~\citep{Trulsson_2012}, 
which can be thought of as the length scale of the particle surface roughness;
the leading terms we use for normal and tangential displacements then behave
as $1/(h^{(i,j)}+\delta)$ and $\log(1/(h^{(i,j)}+\delta))$. 
In the simulations, we set $\delta = 10^{-3}$,
giving large enough resistance for the squeeze mode
compared to the typical hydrodynamic force.
This value is typical for non-Brownian particles~\citep{Davis_1992, Blanc_2011a}.
The detailed expressions for the resistance matrices
$\bm{R}_{\mathrm{Lub}}$ and $\bm{R}'_{\mathrm{Lub}}$
are given in \appref{app:resistance_matrices}. 

\subsubsection{Contact model}
\label{sec:contact_model}

There are several models that one can use to describe frictional contacts
between particles.
We use a stick/slide friction model employing springs and dashpots
that is commonly used in granular physics~\citep{Cundall_1979, Luding_2008}.
The normal and tangential components of the force and the torque
for particles having radii $a_i$ and $a_j$ are obtained as
\begin{equation}
  \label{eq:contact_model}
  \begin{split}
    \bm{F}_{\mathrm{C,nor}}^{(i,j)} & = k_{\mathrm{n}} h^{(i,j)}
    \bm{n}_{ij} + \gamma_{\mathrm{n}}  \bm{U}_{\mathrm{n}}^{(i,j)},  \\
    \bm{F}_{\mathrm{C,tan}}^{(i,j)} & = k_{\mathrm{t}}
    \bm{\xi}^{(i,j)}, \\
    \bm{T}_{\mathrm{C}}^{(i,j)} & = a_i \bm{n}_{ij} \times
    \bm{F}_{\mathrm{C,tan}}^{(i,j)},
  \end{split}
\end{equation}
and fulfill Coulomb's friction law
$\bigl|\bm{F}_{\mathrm{C,tan}}^{(i,j)} \bigr| \leq \mu
\bigl|\bm{F}_{\mathrm{C,nor}}^{(i,j)}\bigr|$.
In the above expressions,
$k_{\mathrm{n}}$ and $k_{\mathrm{t}}$ are 
the normal and tangential spring constants, respectively,
and $\gamma_{\mathrm{n}}$ is the damping constant.
The normal and tangential velocities are
$\bm{U}_{\mathrm{n}}^{(i,j)} 
\equiv \bm{n}_{ij}\bm{n}_{ij}\cdot (\bm{U}^{(j)} - \bm{U}^{(i)})$ 
and $\bm{U}_{\mathrm{t}}^{(i,j)} \equiv 
(\id -\bm{n}_{ij}\bm{n}_{ij}) \cdot [\bm{U}^{(j)} -\bm{U}^{(i)} 
  - ( a_i \bm{\Omega}^{(i)} + a_j \bm{\Omega}^{(j)} )\times \bm{n}_{ij}]$.
Finally, the quantity $\bm{\xi}^{(i,j)}$ is 
the tangential spring stretch. 
This contact model could be made more general, but at the price of
numerical difficulties; see~\appref{app:contact_model_full}.


The computation of the tangential spring stretch $\bm{\xi}^{(i,j)}$,
described in the following, requires some care, 
as we have to impose Coulomb's law.
We apply an algorithm described in~\citet{Luding_2008}.
At the time $t_{0}$ at which the contact ${(i,j)}$ is created, 
we set an unstretched tangential spring
$ \bm{\xi}^{(i,j)}(t_{0}) = \bm{0} $.
At any further time step $t$ in the simulation, the tangential stretch
$\bm{\xi}^{(i,j)}(t)$ is incremented according to the value of a ``test'' force 
$ \bm{F}_{\mathrm{C,tan}}^{\prime(i,j)}(t+\mathrm{d}t) =
 k_{\mathrm{t}}\bm{\xi}^{\prime(i,j)}(t+\mathrm{d}t)$ 
with 
$\bm{\xi}^{\prime(i,j)}(t+\mathrm{d}t) = \bm{\xi}^{(i,j)}(t) +
\bm{U}_{\mathrm{t}}^{(i,j)}(t) \mathrm{d}t$.
If $\bigl|\bm{F}_{\mathrm{C,tan}}^{\prime(i,j)}(t+\mathrm{d}t)\bigr| 
\leq \mu \bigl|\bm{F}_{\mathrm{C,nor}}^{(i,j)}(t+\mathrm{d}t)\bigr|$, 
the contact is in a static friction state
and we update the spring stretch and force as
\begin{equation}
  \begin{split}
    & \bm{\xi}^{(i,j)}(t+\mathrm{d}t)  =
    \bm{\xi}^{\prime(i,j)}(t+\mathrm{d}t), \\
    &\bm{F}_{\mathrm{C,tan}}^{(i,j)}(t+\mathrm{d}t)  =
    \bm{F}_{\mathrm{C,tan}}^{\prime(i,j)}(t+\mathrm{d}t).
  \end{split}
\end{equation}
However, if 
$\bigl|\bm{F}_{\mathrm{C,tan}}^{\prime(i,j)}(t+\mathrm{d}t)\bigr| 
> \mu \bigl|\bm{F}_{\mathrm{C,nor}}^{(i,j)}(t+\mathrm{d}t)\bigr|$, 
the contact is in a sliding state
and the spring and force are updated as
\begin{equation}
  \begin{split}
    &\bm{\xi}^{(i,j)}(t+\mathrm{d}t) = 
    \frac{\mu}{k_{\mathrm{t}}}
    \bigl| \bm{F}_{\mathrm{C,nor}}^{(i,j)}(t+\mathrm{d} t) \bigr| 
    \tanVec
    , \\
    &\bm{F}_{\mathrm{C,tan}}^{(i,j)}(t+\mathrm{d}t)=
    k_{\mathrm{t}} \bm{\xi}^{(i,j)}(t+\mathrm{d}t),
  \end{split}
\end{equation}
where the direction $\tanVec$ is the same as the one for the test force,
i.e.,
$ \tanVec \equiv \bm{F}_{\mathrm{C,tan}}^{\prime(i,j)} (t+\mathrm{d}t)
/ \bigl|\bm{F}_{\mathrm{C,tan}}^{\prime(i,j)}(t+\mathrm{d} t)\bigr| $.
In this case, Coulomb's law is not violated, as
$ \bigl|\bm{F}_{\mathrm{C,tan}}^{(i,j)}(t+\mathrm{d}t)\bigr| = 
\mu \bigl|\bm{F}_{\mathrm{C,nor}}^{(i,j)}(t+\mathrm{d} t)\bigr|$.

\subsection{Other interactions, full models}
\label{sec:repulsion}

\subsubsection{Shear rate dependence: why Stokes hydrodynamics plus hard spheres is not enough}
\label{sec:shearrate_dep}

The shear-rate dependence of a suspension requires
the existence of a time scale distinct from the inverse shear rate.
A well known case is the one of Brownian suspensions, 
where the inverse shear rate competes with the Brownian diffusion time 
to give a  shear-rate dependent rheology. 
This competition is adequately
measured by a non-dimensionalized shear rate, the Péclet number, which
is the ratio of the two time scales: 
$\PeNum \equiv 6 \pi \eta_0 a^3 \dot{\gamma}/k_{\mathrm{B}}T$.
This quantity can also be thought of as a ratio of two force scales:
the typical hydrodynamic force and the typical Brownian force.
Thus, a shear-rate dependent rheology generically requires at least one other force scale
besides the hydrodynamic one.


Such a time scale does not exist 
for non-Brownian suspensions of hard spheres in the Stokes regime.
The reason is straightforward:
the hard sphere force has no typical value, 
and a hard-sphere contact between two particles can withstand any applied load.
Therefore there is nothing with which to compare the hydrodynamic force.
This means that the solutions of the force/torque balance equations~\eqref{eq:fb}
(i.e., the particle trajectories) 
are independent of the shear rate, 
leading to a shear-rate independent rheology.
Notice that this conclusion holds whether the hard spheres 
are frictionless or frictional: 
Coulomb's law does not introduce a force scale.
There must be another force scale
in the system to provide the shear-rate dependence.


An important point in our model is that, even if we try to mimic a
hard sphere suspension, our contact forces actually are elastic
forces, thus bringing a natural force scale
$k_{\mathrm{n}}a$. 
We preserve the shear-rate independent hard sphere behavior by
selecting the particle stiffness $k_{\mathrm{n}}$ such that the
non-dimensional number based on the stiffness $6 \pi \eta_0 a\dot{\gamma}/k_{\mathrm{n}}$ 
remains much smaller than unity;
i.e.,
we stay in the asymptotic regime of nearly hard particles. 
We describe the techniques we use to achieve this in our simulation
in~\appref{app:contact_model_full}.

\subsubsection{A minimal model: critical-load friction}

The first model we consider is an intentionally simplistic one,
arguably the simplest model with an additional force scale 
besides the hydrodynamic force.
Simplicity is the main motivation for this model, 
as it enables the clear physical discussion one can expect from a minimal model. 


In this model (which we call Critical Load Model, CLM), 
we introduce an extra force scale in the friction law itself,
namely a threshold normal load $F_{\mathrm{CL}}$ to activate friction.
When the normal load is smaller than this threshold, particles
interact as frictionless hard spheres.
When the load is beyond the threshold, friction is activated.
Overall, the friction law reads:
\begin{equation}
  \bigl|\bm{F}_{\mathrm{C,tan}}^{(i,j)}\bigr| 
  \leq
  \begin{cases}
    \mu \Bigl(\bigl|\bm{F}_{\mathrm{C,nor}}^{(i,j)}\bigr|-F_{\mathrm{CL}}\Bigr) 
    & \text{for $\bigl|\bm{F}_{\mathrm{C,nor}}^{(i,j)}\bigr| \geq F_{\mathrm{CL}}$}, \\
    0 
    & \text{otherwise}.
  \end{cases}
\end{equation}
We may consider the CLM as the short Debye length limit 
of the electrostatic repulsion model presented below (\secref{sec:es_repulsion_model}).

\subsubsection{Electrostatic repulsion model}
\label{sec:es_repulsion_model}

An electrostatic repulsion is a simple and plausible interaction.
Many experimental systems have
such a force, as it stabilizes the suspension.
Thus, we will consider an Electrostatic Repulsion Model (ERM)
including this ingredient in the simplest form of 
a repulsive double-layer electrostatic force
$\bm{F}_{\mathrm{R}}^{(i,j)}$~\citep{Israelachvili_2011}.
If the Debye length $\kappa^{-1}$ is small compared to the radius of the
particles, the approximate form
\begin{equation}
  \label{eq:repulsive_force}
  \bm{F}_{\mathrm{R}}^{(i,j)}(h) \simeq
  \begin{cases}
    -2 F_{\mathrm{ER}} a_i a_j/(a_i+a_j) 
    e^{- \kappa (r^{(i,j)}- a_i-a_j )} \bm{n}_{ij} 
    & \text{if $h^{(i,j)}\geq 0$,} \\
    -2 F_{\mathrm{ER}}  a_i a_j/(a_i+a_j) \bm{n}_{ij} 
    & \text{if $h^{(i,j)}<0$,}
  \end{cases}
\end{equation}
can be used.
In the simulations, we set $\kappa^{-1} = 0.05 a$, meaning a Debye
length in the \SI{0.1}{\micro\meter} range for particles of a few
\si{\micro\meter}; see \citet{Mewis_2011}.

\subsection{Stress tensor and bulk rheology}

The mechanical stress applied to the suspension arises from the
several interactions included in the model: particles disturb the flow,
creating hydrodynamic stresses, and they develop force chains via contacts
(and/or electrostatic repulsion for the ERM).
%


%
The hydrodynamic stresslets acting on the particles 
are given by~\citep{Batchelor_1970a,Brady_1988}
\begin{equation}
  \bm{S}_{\mathrm{H}} = 
  - 
  \bigl(\bm{R}_{\mathrm{Stokes}}^{S}+ \bm{R}_{\mathrm{Lub}}^{S} \bigr)
  \cdot 
  \begin{pmatrix}
    \bm{U} - \bm{U}^{\infty} \\
    \bm{\Omega} - \bm{\Omega}^{\infty} 
    \end{pmatrix}
    +
    \bm{R}_{\mathrm{Lub}}^{\prime S}
    :\bm{E}^{\infty}
  ,
\end{equation}
where 
$\bm{S}_{\mathrm{H}} \equiv 
\bigl(\bm{S}_{\mathrm{H}}^{(1)}, \dotsc, \bm{S}_{\mathrm{H}}^{(n)} \bigr)$
and the matrices $\bm{R}_{\mathrm{Lub}}^{S}$ 
and $\bm{R}_{\mathrm{Lub}}^{\prime S}$ contain 
leading terms of the lubrication resistances~\citep{Jeffrey_1984,Jeffrey_1992}
(see \appref{app:resistance_matrices}) in a manner consistent with the
hydrodynamic forces considered in~\secref{sec:hydrodynamic_intereactions}.


The stress due to the contact or repulsive force between particles $i$ and $j$
is simply written as
\begin{equation}
  \bm{S}_{\mathrm{C}}^{(i,j)} = 
  (\bm{r}^{(j)}-\bm{r}^{(i)}) 
  \bm{F}^{(i,j)}_{\mathrm{C}}
  \text{ or }
  \bm{S}_{\mathrm{R}}^{(i,j)} = 
  (\bm{r}^{(j)}-\bm{r}^{(i)}) 
  \bm{F}^{(i,j)}_{\mathrm{R}},
\end{equation}
respectively.
%


The bulk stress, in which the isotropic part of the fluid pressure is omitted,
is the sum of the above contributions:
\begin{equation}
  \bm{\Sigma} \equiv
  2 \eta_0 \bm{E}^{\infty} + 
  \frac{1}{V}
  \biggl(
    \sum_{i}
    \bm{S}_{\mathrm{H}}^{(i)}+
    \sum_{i > j}
    \bm{S}_{\mathrm{C}}^{(i,j)}
    +
    \sum_{i > j}
    \bm{S}_{\mathrm{R}}^{(i,j)}
    \biggr),
\end{equation}
where $V$ is the volume of the simulation box 
(the electrostatic stress $\bm{S}_{\mathrm{R}}$ appears only for the ERM).
Shear stress $\sigma$,
normal stress differences $N_1$ and $N_2$,
and particle pressure $\Pi$~\citep{Yurkovetsky_2008} are defined as
$\sigma \equiv \Sigma_{12}$,
$N_1 \equiv \Sigma_{11} - \Sigma_{22}$,
$N_2 \equiv \Sigma_{22} - \Sigma_{33}$,
and 
$\Pi \equiv -(\Sigma_{11} + \Sigma_{22} + \Sigma_{33})/3$,
respectively.
The relative viscosity $\eta_{\mathrm{r}}$ is given by
$\eta_{\mathrm{r}} \equiv \sigma / \eta_0 \dot{\gamma}$.

\subsection{Additional points in the simulation model}
\label{sec:simulation_details}

\paragraph{Bidispersity:}

We consider a suspension of bidisperse frictional hard spheres
immersed in a Newtonian fluid with viscosity $\eta_0$. 
The bidispersity is introduced to hinder the formation of the ordered
phase observed for dense monodisperse suspensions under shear.
An effective choice for the size ratio is $a_2/a_1 =1.4$ (where $a_1=a$), 
with the two populations occupying equal volumes; 
i.e., $\phi_1 = \phi_2$. 
Other effective size ratios are possible,
as is polydispersity, with the same qualitative effects.
The choice of size ratio $1:1.4$ is however the most common one in the
granular matter literature, where bidispersity is used to avoid crystallization
(see for example~\citep{OHern_2003}).
Indeed,
with a smaller size ratio $1:1.2$,
slow and weak strain thinning due to ordering
is seen in the low viscosity state.

\paragraph{Dimensionless shear rate:}

We discussed in \secref{sec:shearrate_dep} that 
another force scale besides the hydrodynamic one
is required to yield shear-rate dependence.
In CLM, 
the threshold value gives the force scale $F^{\ast} = F_{\mathrm{CL}}$,
and
in ERM,
the force at contact gives $F^{\ast} = F_{\mathrm{ER}}$.
Therefore, the shear rate dependence is given by the ratio
$\dot{\gamma}/\dot{\gamma}_0$, with $\dot{\gamma}_0 \equiv F^{\ast}/
6\pi \eta_0 a^2 $.

\paragraph{Periodic boundary conditions and fixed-volume simulation:}

Rheology is a bulk property.
If solid walls are used for the boundaries of the system,
we need a large system size to reduce the influence of walls.
We can avoid the use of solid walls in a simulation by using periodic
boundary conditions.
For the strain-controlled simple shear,
we use the Lees-Edwards~\citep{Lees_1972} boundary condition.


Particle migration never develops under these periodic boundary conditions.
Although some experimental observations
suggest that  global migration 
may be a cause of shear thickening~\citep{Fall_2010},
our simulation is free of this effect.
%


There is no way for the system to dilate with these periodic boundary
conditions.
Such a fixed volume condition is expected in most rheology measurements.
For shear thickening fluids, however, 
the open edges may have some influence~\citep{Brown_2012}.
In granular physics,
systems are often sheared under a given normal stress,
hence the volume is  not fixed~\citep{Boyer_2011}.

\section{Results}

This section presents the results from the simulation of the two
models, the ``minimal'' critical load model (CLM) and the electrostatic
repulsion model (ERM).
In the following subsections, 
whenever possible, the two models are treated at the same time in the text,
and we will show data plots for the CLM.
However, when the two models give slightly different results, 
the results of the CLM will be described first, 
as it allows a simple and clear understanding of the underlying physics,
and the more realistic model ERM will be described afterwards.
The differences are essentially in the low shear rate limit
for the ERM (existence of a shear-thinning regime), due to the repulsive potential.


Although friction is the key factor in this work,
few experimental estimates are available for the friction coefficient.
We select a friction coefficient $\mu=1$
(which is comparable to the measurements of~\citet{Fernandez_2013} for
$\sim \SI{10}{\micro\meter}$ quartz particles coated by polymer brushes)
for most of the simulations, except when the dependence on $\mu$ is
specifically investigated.


In the data plots shown in this section, the error bars represent the
standard deviation, which we define for an observable $A(t)$ as
$\sqrt{\langle A^2 \rangle - \langle A \rangle^2}$, where $\langle
\cdot \rangle$ is the time (strain) average $T^{-1}\int_0^T \cdot \,
\mathrm{d}t$ over the simulated units of strain. (We perform
simulations over $50$ strain units, hence $T=50$ except when the time
averages are performed over subsets of the whole simulation in the
case of intermittent data around DST).

Rheological data are plotted
versus shear rates and shear stresses non-dimensionalized by the natural
shear rate $\dot\gamma_0 = F^{\ast}/ 6\pi \eta_0 a^2 $ and stress
$\eta_0\dot\gamma_0$, respectively, where $\eta_0$ is the viscosity of the suspending fluid.

\subsection{Frictionless and frictional rheologies}
\label{sec:fric_fricless}

In the CLM, due to the threshold force, 
the friction between grains is absent at low shear rates
and activated at high shear rates. 
Because of this, we expect that the low shear-rate limit for concentrated
suspensions will have 
a rheology $\eta(\phi, \dot\gamma \to 0)$ typical of a system 
close to the jamming transition for frictionless particles, 
while the high shear-rate limit shows 
a rheology $\eta(\phi, \dot\gamma\to \infty)$ typical of a system 
close to the jamming transition for particles with a friction coefficient $\mu = 1$.

\begin{figure}[hbt]
\centering
\includegraphics[width=0.9\textwidth]{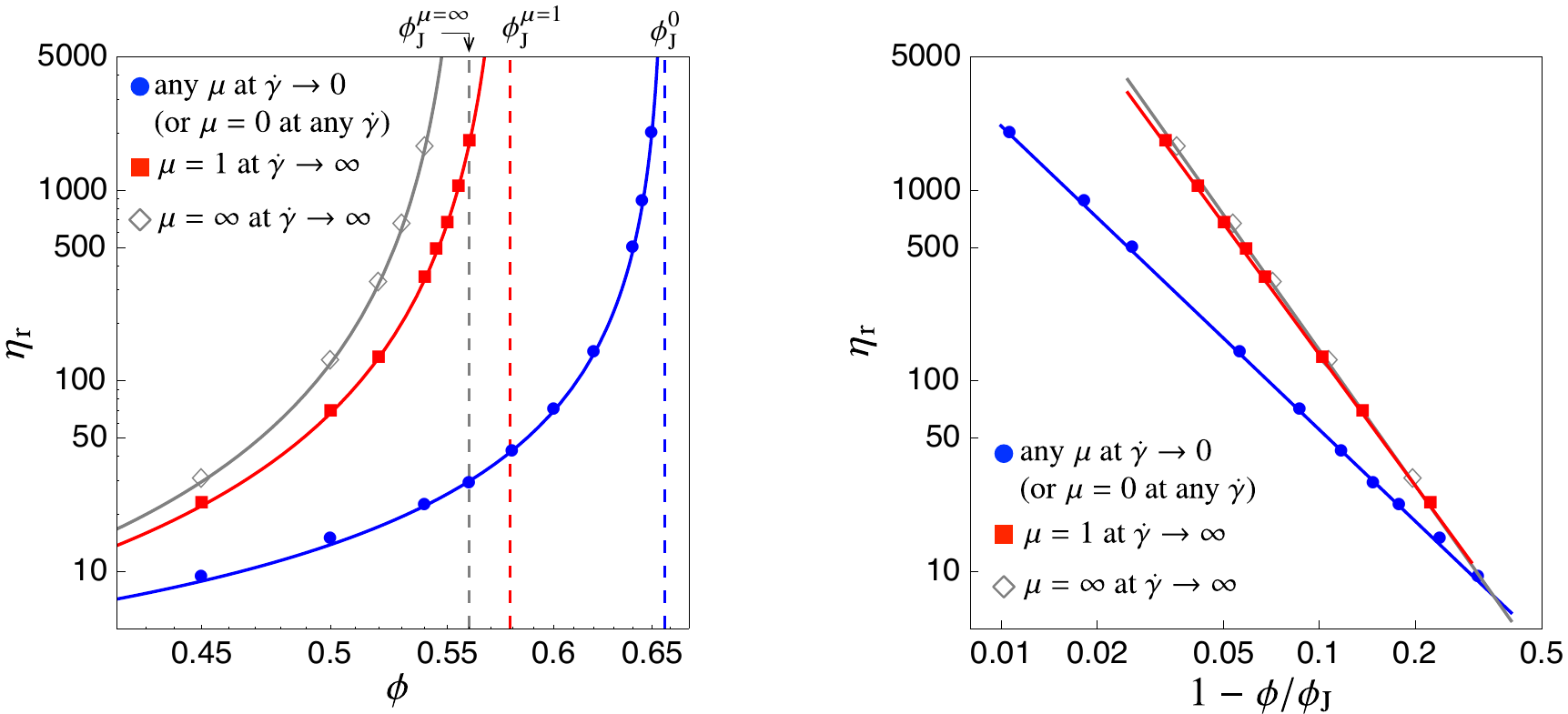}
\caption{
Relative viscosity $\eta_{\mathrm{r}}$
as a function of the volume fraction $\phi$
in the two limits $\dot \gamma \to 0$ and $\dot \gamma \to \infty$ (left).
The $\dot \gamma \to 0$ viscosity (blue circles) is independent 
of the friction coefficient $\mu$ 
as the friction is not activated at low stresses,
which leads to a relatively lower viscosity diverging 
at a higher volume fraction $\phiJam$ 
(which is the jamming point for frictionless systems).
The $\dot \gamma \to \infty$ viscosity however directly depends on $\mu$,
as is seen from the difference 
between $\mu=1$ (red squares) and $\mu=\infty$ (gray diamonds) plots.
In particular, the jamming volume fraction decreases with increasing $\mu$.
We fit our data with power laws 
$\eta_{\mathrm{r}} = C (1 - \phi/\phi_{\mathrm{J}})^{-\lambda}$ (right).
The best fitting parameters are 
$(\phiJam, \lambda^{0}, C^{0}) \approx (0.66, 1.6, 1.40)$,
$(\phi_{\mathrm{J}}^{\mu=1}, \lambda^{\mu=1}, C^{\mu=1}) \approx (0.58, 2.3, 0.71)$,
and 
$(\phi_{\mathrm{J}}^{\mu=\infty},\lambda^{\mu=\infty}, C^{\mu=\infty})
\approx (0.56, 2.4, 0.63)$.}
\label{fig:two_rheologies}
\end{figure}

These two limiting viscosities are shown in~\figref{fig:two_rheologies},
where we also show the high shear-rate behavior for the infinite
friction case ($\mu=\infty$) for reference.
Each diverges at a different volume fraction,
thus friction shifts the jamming point~\citep{Otsuki_2011}.
We fit our data with power law divergences $\eta \propto
C(1-\phi/\phi_{\mathrm{J}})^{-\lambda}$, with parameters
$(\phi_{\mathrm{J}}, \lambda, C)$ as detailed in the caption of~\figref{fig:two_rheologies}.


For the ERM, the situation is the same at high shear rates, 
but differs at low shear rates. 
While friction is not felt for $ \dot \gamma \to 0 $, 
as particles do not contact, 
the finite range of the repulsive potential creates a shear thinning behavior 
from which we could not obtain the low shear-rate limit $\eta(\phi, \dot\gamma\to 0)$. 
The shear thinning comes from the fact that the system behaves
essentially as soft particles at low shear stresses, with an apparent
size that includes the hard sphere and a part of the surrounding soft
repulsive potential.
Indeed, a simple force balance argument at the scale of a particle
says that if a particle is subject to a driving shear force $\sigma a^2$,
the minimum gap $h_{\mathrm{min}}$ between this particle and its
neighbors will be such that
$F_{\mathrm{R}}(h_{\mathrm{min}}) \sim \sigma a^2$
(with the limitation that it must respect the geometrical
constraint, i.e.,
$h_{\mathrm{min}} < \bigl(1-\phi/\phi_{\mathrm{J}}^{0}\bigr)^{1/3}$).
Then, at this shear stress the minimum center-to-center distance
between two particles is $2a(1+h_{\mathrm{min}}(\sigma)/2)$, which means
that the particles have an apparent radius larger than the one of
their hard core.
The jamming transition of these effectively bigger frictionless particles 
is at an apparent packing fraction (that is, based on the apparent diameter) 
$\phi' = \phi_{\mathrm{J}}^{\prime 0} \approx 0.66$, 
which corresponds to an actual packing fraction $\phi$ significantly lower 
than $\phi_{\mathrm{J}}^{\prime 0}$.
Because of this low volume fraction jamming transition at $\dot \gamma = 0$, 
the viscosity is larger than at finite (but small) $\dot\gamma$, 
leading to the shear thinning we observe.
Such thinning is actually known to occur due to repulsive forces
in charge stabilized suspensions~\citep{Krieger_1972,Maranzano_2001a}.

\subsection{Shear thickening, continuous and discontinuous}

We can switch from one rheology to the other by varying the shear rate.
Physically, the transmitted stress increases as the shear rate increases, 
which triggers the formation of frictional contacts between particles.
Thus, by increasing the shear rate, the viscosity interpolates between
the frictionless and frictional rheology curves, 
which means we can observe shear thickening.
All this should be a natural consequence 
of the existence of two distinct rheologies at $\dot\gamma = 0$ and $\dot\gamma = \infty$. 
What we cannot  anticipate \textit{a priori} is the way in which the
system switches from the low viscosity state to the high viscosity one:
do we observe a Continuous Shear Thickening (CST) 
or a Discontinuous Shear Thickening (DST)?

\begin{figure}[htb]
\centering
\includegraphics[width=0.95\textwidth]{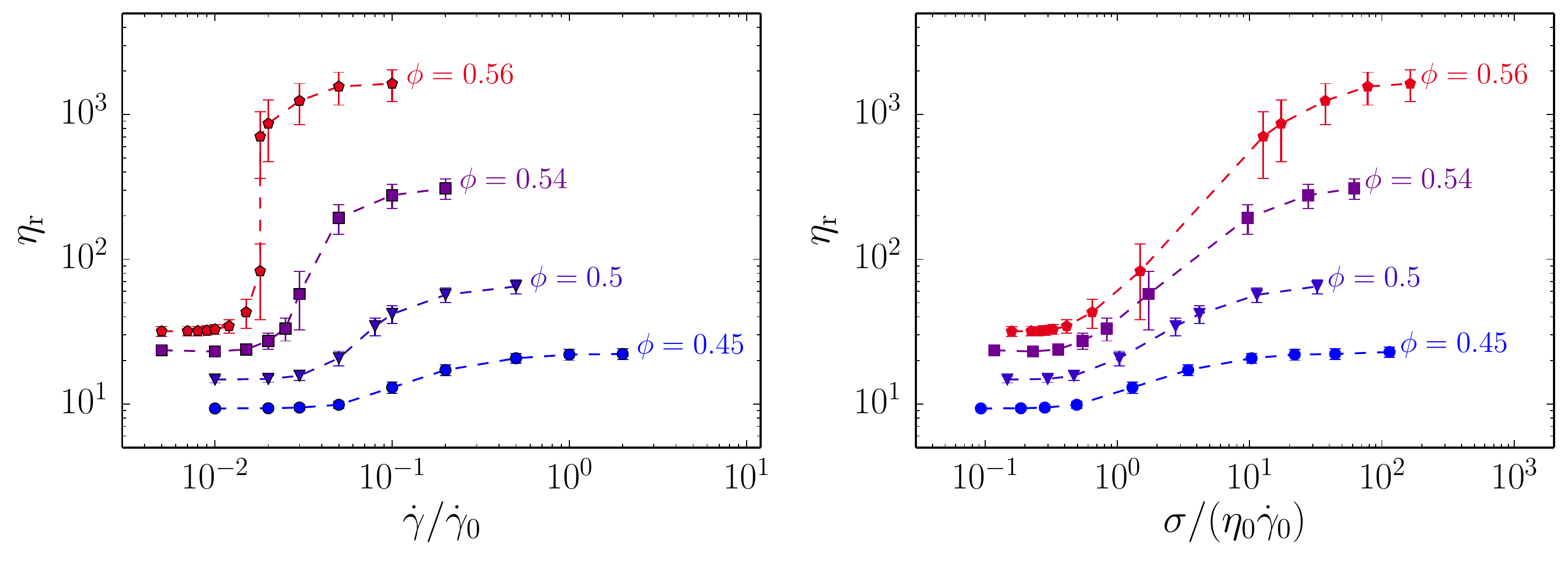}
\caption{
Shear rate $\dot{\gamma}$ (left) and shear stress $\sigma$ (right) dependences of 
the relative viscosity $\eta_{\mathrm{r}}$ for the critical load model (CLM), 
with friction coefficient $\mu=1$, for volume fractions $0.45\leq \phi \leq 0.56$. 
The system size is $N=1000$.
}\label{fig:visc_threshold}
\end{figure}

\begin{figure}[htb]
  \centering
  \includegraphics[width=0.45\textwidth]{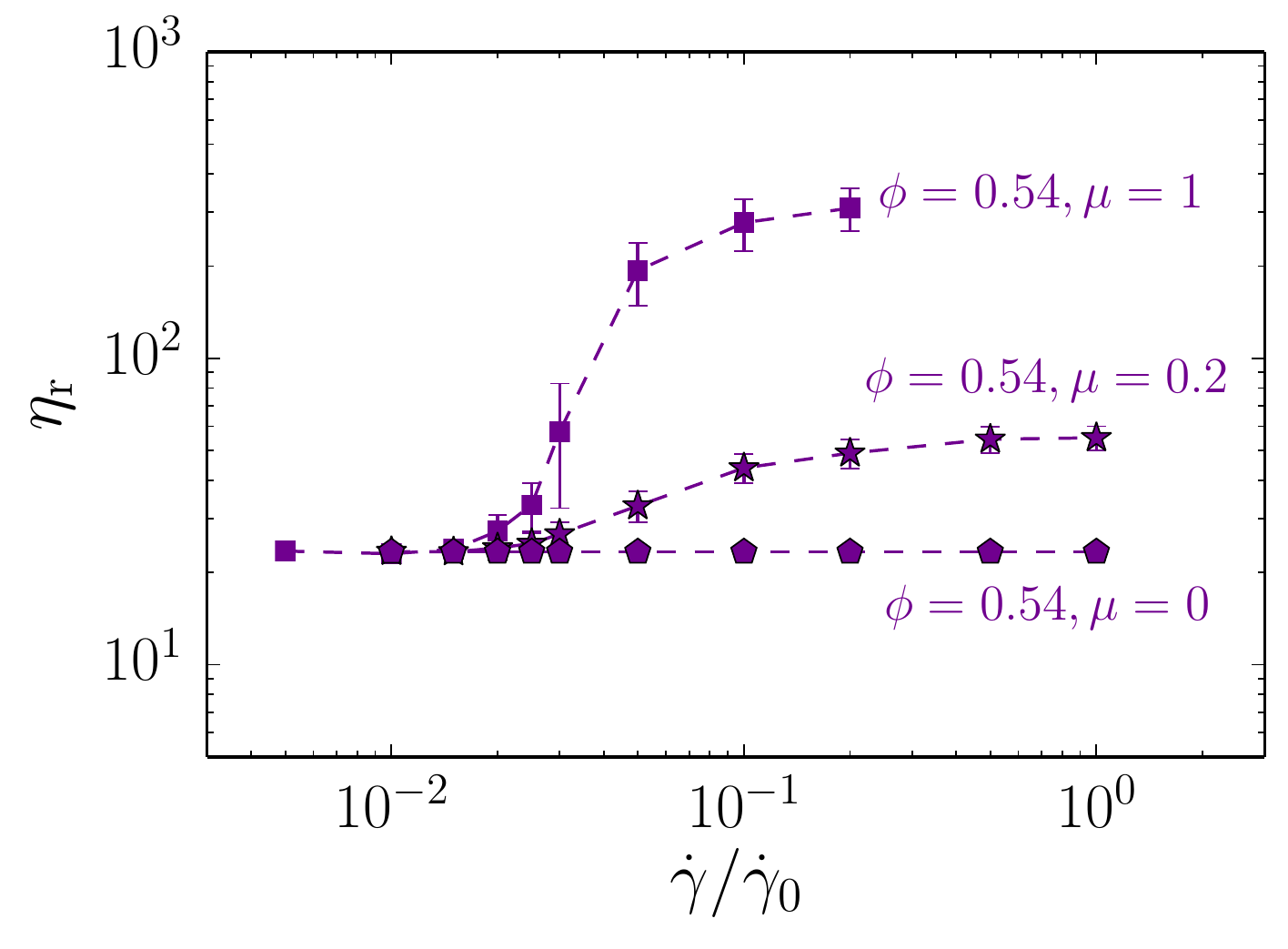}
  \caption{
    Effect of the friction coefficient $\mu$ 
    on the shear rate dependence of the relative viscosity $\eta_{\mathrm{r}}(\dot{\gamma})$
    in the CLM.
    The friction is essential to the shear thickening, 
    as is illustrated by the reduction of the effect for $\mu=0.2$ 
    and its complete suppression for $\mu=0$.
  }\label{fig:visc_mudep}
\end{figure}

\begin{figure}[htb]
\centering
\includegraphics[width=0.95\textwidth]{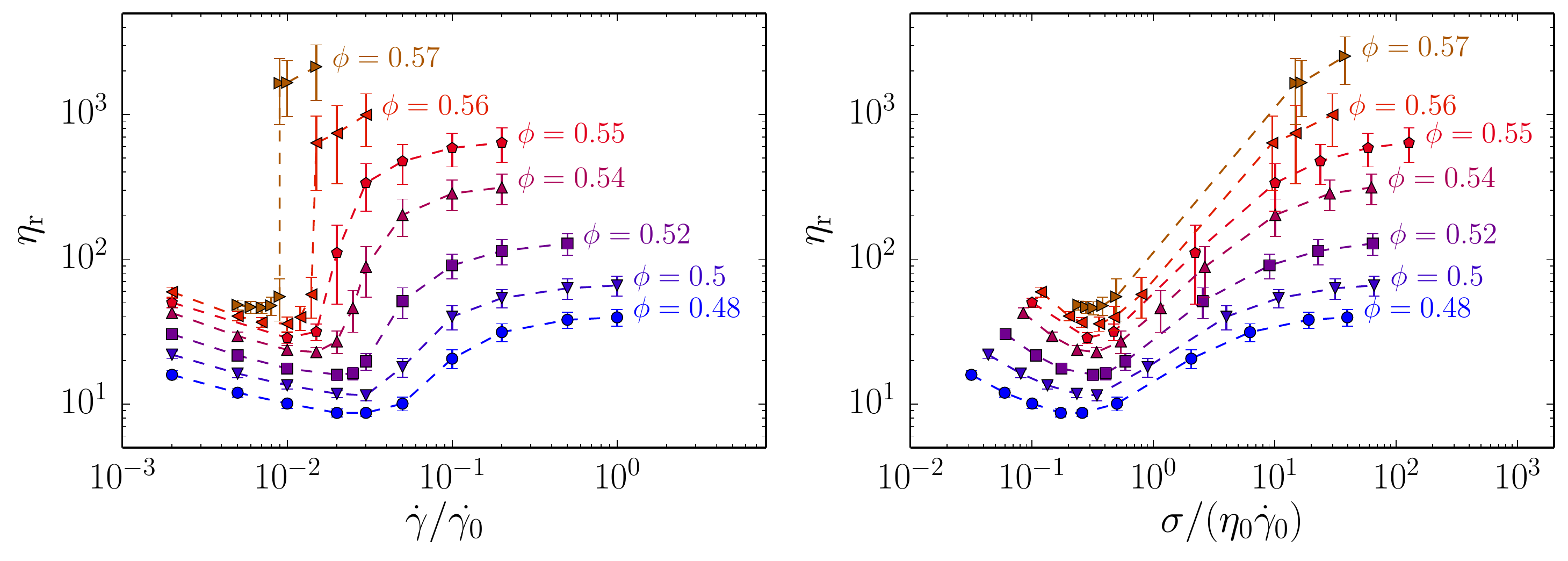} 
\caption{
  Shear rate $\dot{\gamma}$ (left) and shear stress $\sigma$ (right) dependences of 
  the relative viscosity $\eta_{\mathrm{r}}$ 
  for the electrostatic repulsion model (ERM), 
  with friction coefficient $\mu=1$, 
  for volume fractions $0.48\leq \phi \leq 0.57$.
  The system size is $N=512$.
}\label{fig:visc_repulsion}
\end{figure}


The shear rate dependence of the viscosity, 
shown in~\figref{fig:visc_threshold} for the CLM, 
demonstrates the existence of both CST and DST in our system, 
depending on the volume fraction. 
As in experiments, when $\phi < \phiDST$ the shear thickening is continuous, 
getting steeper and steeper as we approach $\phiDST$, 
at which point it becomes discontinuous 
and keeps this behavior for $ \phi > \phiDST$ and  up to $\phiJam$.
Note that there appears to be a real discontinuity in our data for
these volume fractions: 
the time series of the viscosity show an intermittent behavior 
switching between two states, 
which we address in~\secref{sec:discontinuity_hysteresis}.
In~\figref{fig:visc_threshold}, the intermittent data are split between
low and high viscosity states: two points that correspond to a separate time
average for each of the two states appear at the same shear rate.


When plotted against stress in~\figref{fig:visc_threshold}, 
the viscosity curves show another interesting feature, 
namely that the onset of shear thickening occurs at a stress
$ \sigma_{\mathrm{ST}}/(\eta_0 \dot{\gamma}_0) \approx 0.3 $
that is roughly independent of the volume fraction,
as is observed in many experiments~%
\citep{Frith_1996, Bender_1996, Maranzano_2001, Maranzano_2001a,
  Lootens_2005, Fall_2010, Larsen_2010, Brown_2012, Brown_2014}.
Similarly, the transition towards the
shear-rate independent plateau at high viscosity also occurs at
a stress independent of $\phi$.
The stress range over which thickening occurs remains constant from a
mild shear thickening to a marked DST, as expected from a simple
balance argument between the driving stress $\sigma = \eta \dot\gamma$
and the stress required to create a frictional contact 
$\sim F_{\mathrm{CL}} / a^2$ (for the CLM) with the area $a^2$ meaning that
we require almost every particle to have a contact.
It implies the thickening should roughly take place around 
$\sigma \approx F_{\mathrm{CL}}/a^2 = 6\pi \eta_0 \dot{\gamma}_0$, 
which gives $\sigma/(\eta_0 \dot{\gamma}_0)\approx 6\pi$.
This value falls at the upper end of the thickening regime as shown
in~\figref{fig:visc_threshold}, which is consistent with our assumption
that at this stress scale every particle should have contacts with its neighbors, hence the
system is in the frictional rheology state.
The inverse argument applied to the onset stress
$\sigma_{\mathrm{ST}}$ finds that the area $A_{\mathrm{ST}}$ 
over which one unit of critical load applies is
$\sigma_{\mathrm{ST}} = F_{\mathrm{CL}}/A_{\mathrm{ST}}$, 
with $A_{\mathrm{ST}} \approx (7a)^2$.
That means that at the onset of shear thickening, the contact chains
must be sparsely distributed.


Friction is an essential ingredient for the shear thickening to occur,
as is illustrated in~\figref{fig:visc_mudep}, 
which shows the effect of a reduction of the friction coefficient 
on the $\eta_{\mathrm{r}}(\dot\gamma)$ curve in the CST regime. 
If the friction is removed ($\mu=0$), the effect is completely suppressed 
(which is expected, as the threshold force scale $F_{\mathrm{CL}}$
disappears from the equations of motion for $\mu=0$,
and the simulation is then rigorously the same for all $\dot\gamma$).
If the friction is only reduced ($\mu=0.2$),
the shear thickening is milder than with $\mu=1$ friction at the same volume fraction.
In that case, the CST however becomes more pronounced and turns to DST 
when one increases the volume fraction, just as for the $\mu=1$ case. 


The ERM shown in~\figref{fig:visc_repulsion}
behaves in a similar manner,
only adding a shear thinning regime at low shear rates
as discussed in the last paragraph of \secref{sec:fric_fricless}.

\subsection{Discontinuity and hysteresis}
\label{sec:discontinuity_hysteresis}

The existence of two distinct flowing states suggests that 
one might expect to observe a hysteresis associated with DST. 
It is thus remarkable that, 
aside from suspensions that are hysteretic at the microscopic level 
(with attractive interactions, for example, since once particles are
aggregated under flow they stay aggregated upon cessation of the flow),
there are very few experimental reports of hysteresis in systems showing DST 
(one of the best examples is provided by~\citet{Bender_1996}).


\begin{figure}[htb]
\centering
\includegraphics[width=0.95\textwidth]{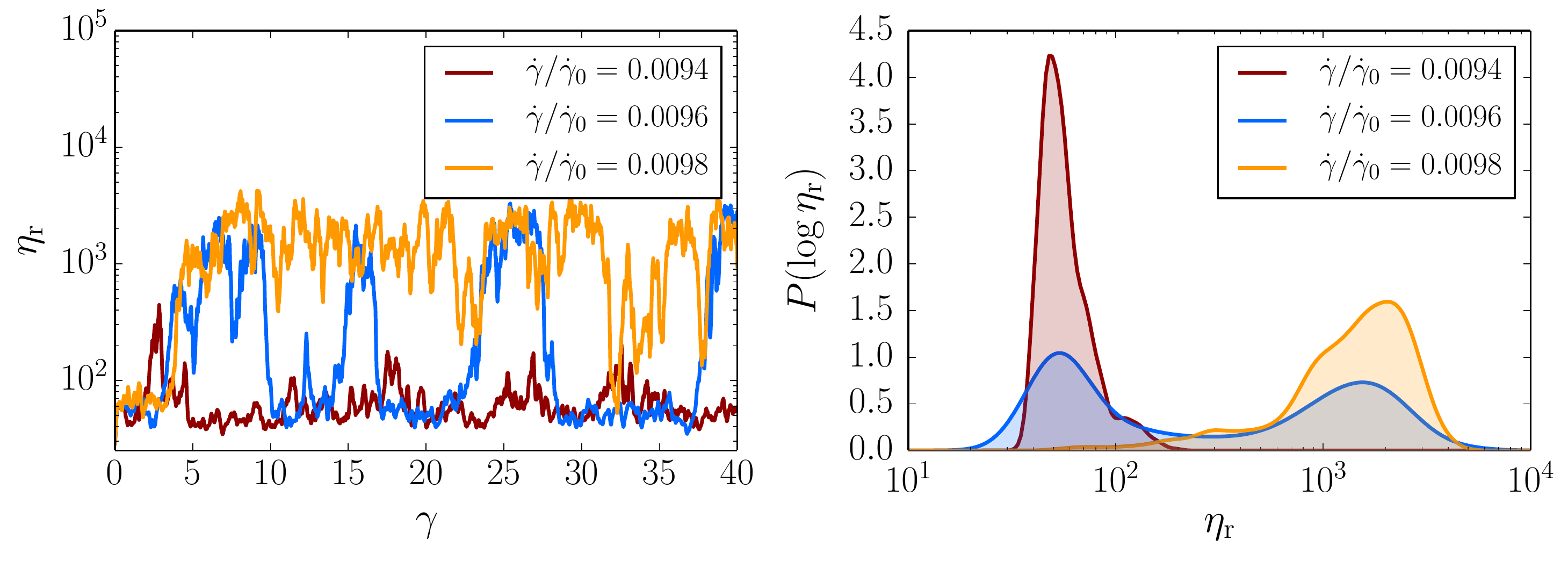}
\caption{
  Strain (or dimensionless time) series of the shear stress close to the DST (left), 
  and the corresponding histograms (right).
  The simulations are performed at $\phi=0.57$ with
  the ERM, with shear rates $\dot\gamma/\dot\gamma_0=0.0094$ (just below the DST region),
  $\dot\gamma/\dot\gamma_0=0.0096$ (in the DST region), and $\dot\gamma/\dot\gamma_0=0.0098$
  (just above the DST region). 
  Data at $\dot\gamma/\dot\gamma_0=0.0096$ show a superposition of the
  two states.
  This is very similar to what is seen in experiments~\citep{Boersma_1991}.}
\label{fig:stress_histograms_repulsion}
\end{figure}

What is commonly observed, however, is a switching behavior, 
where the time series of the stress shows that 
the system shares its time between the two states, 
erratically switching in what looks like activated events~%
\citep{Boersma_1991, DHaene_1993, Bender_1996, 
  Lootens_2003, Lootens_2004, Lootens_2005, Hebraud_2005}.
We observe this switching behavior in both models that we studied. 
In~\figref{fig:stress_histograms_repulsion} 
we show time series of the stress and the corresponding
histograms around the DST.
These data show a typical ``coexistence'' behavior: significantly
below and above shear thickening, the system respectively stays in the
low and high viscosity state, but close to DST,
the two states coexist in the same time series, 
the proportion of
time spent in the high viscosity state increasing with $\dot\gamma$.
While it is tempting to use the analogy of DST with an equilibrium
first-order transition to conclude that the switchings are finite-size ``activated'' events,
we need to be careful in doing so,
because intermittency seems to survive for unusually large numbers of particles.
The intermittency is still present in simulations with $N=1000$
particles, and the switching frequency does not seem to decrease
relative to simulations with $N=512$ particles, whereas in a first
order thermodynamic transition, one would expect the switching rate to
be drastically reduced by a doubling of the system size.
In fact, this effect might persist for much larger particle numbers,
as is suggested by the experimentally observed intermittency~%
\citep{Boersma_1991, DHaene_1993, Bender_1996,
  Lootens_2003, Lootens_2004, Lootens_2005, Hebraud_2005}
in systems where $N$ is considerably larger.
The simulations of~\citet{Heussinger_2013} also show intermittency for
systems of several thousands of particles.
Moreover, should this phenomenon disappear in the thermodynamic limit,
the fact that this limit is not observed until there are very large numbers of
particles might be valuable information about the nature of this
out-of-equilibrium phase transition, just as strong finite-size
effects have a profound physical significance near a critical point.


In order to observe hysteresis
one must have a system where the relaxation time $\tau_{\mathrm{relax}}$ 
(the length of the transient) is much smaller than the activation time
$\tau_{\mathrm{act}}$ (the typical time between two switches). 
One then does a proper measurement with a shear rate 
sweep on a time scale $\tau_{\mathrm{sw}}$ such that 
$\tau_{\mathrm{relax}} \ll \tau_{\mathrm{sw}} \ll \tau_{\mathrm{act}}$: 
the first inequality ensures that the system is always in a steady state during the sweep 
and that the measurement is done with sufficient time averaging,
while the second inequality ensures that we are not averaging data over two distinct states.
In our simulations, as can be seen by looking at the time series
in~\figref{fig:stress_histograms_repulsion}, 
a clear separation of time scales is not achieved 
(we have $\tau_{\mathrm{relax}} \lesssim \tau_{\mathrm{act}}$, which
does not permit finding a proper $\tau_{\mathrm{sw}}$) 
and hysteresis cannot be observed unambiguously. 
Many experimental data actually suffer from the same limitations, 
and this might be the reason for the very small number of hysteretic flow curves 
actually reported. 
One exception is the noted work of~\citet{Bender_1996}.

\subsection{Normal stresses}

\begin{figure}[htb]
\centering
\includegraphics[width=0.95\textwidth]{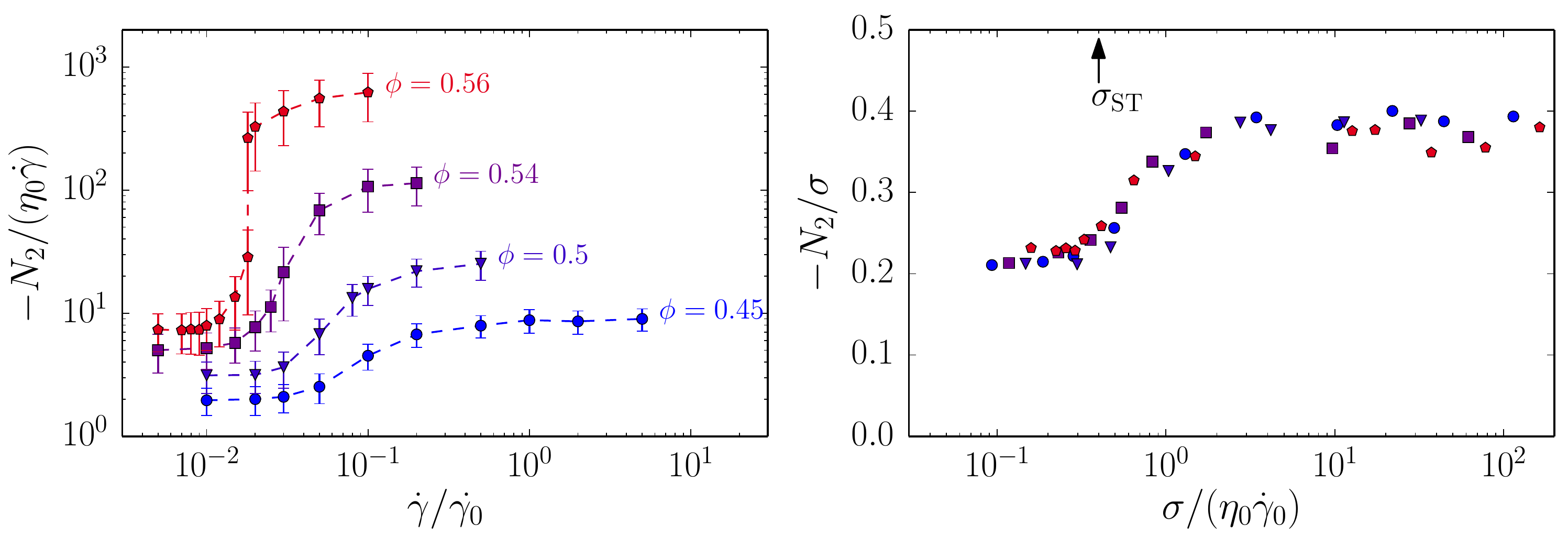} 
 \caption{
 Shear rate dependence of the normalized
second normal stress difference $N_2/\eta_0 \dot\gamma$ (left)
and shear stress dependence of $N_2/\sigma$ (right) for the CLM.
The stress $\sigma_{\mathrm{ST}}$ at which thickening starts is marked
 by an arrow.}
\label{fig:N2_threshold}
\end{figure}

\begin{figure}[htb]
\centering
\includegraphics[width=0.95\textwidth]{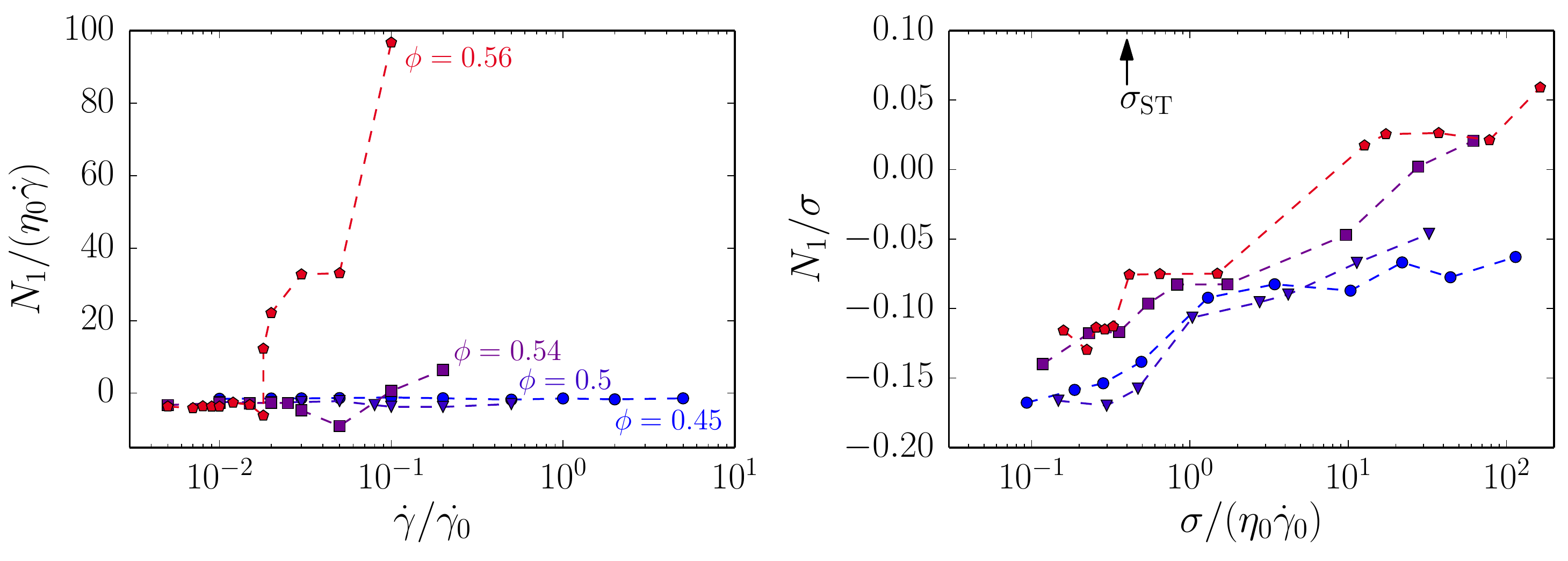} 
 \caption{
 Shear rate dependence of the normalized first normal stress
  difference $N_1/\eta_0 \dot\gamma$ (left) and shear stress
  dependence of $N_1/\sigma$ (right) for the CLM.
}\label{fig:N1_threshold}
\end{figure}

One common rheological characteristic of dense suspensions is 
the appearance of finite normal stress differences. 
Our measurements of the two normal stress differences 
$N_1$ and $N_2$ are shown in~\figref{fig:N1_threshold}
and \figref{fig:N2_threshold} for the CLM.
The ERM behaves very similarly and is not shown.


We obtain for $N_2$ a behavior consistent with most experimental data available:
it is negative, comparable to the shear stress
(and much larger than $N_1$ that we discuss in the next paragraph),
and its behavior is reminiscent of that of the shear stress.
This comes from the fact that most of the stress is transmitted by
forces in the
shear plane (flow-gradient), not along the vorticity direction. 
Indeed the ratio $N_2/\sigma$ is rather insensitive to the stress magnitude,
increasing at most by a factor of two across shear thickening 
while the stresses increase by almost two orders of magnitude. 
What is remarkable is the independence of $N_2/\sigma$ on the volume fraction: 
all volume fractions investigated here collapse on a master curve for both models. 


There has been debate in the community regarding the value (and even
the algebraic sign) of $N_1$.
The first notable feature of our data concerning $N_1$ 
is that it is dominated by the fluctuations.
In one time series, a cursory look indicates that its value fluctuates around zero. 
Long time averaging reveals more structure, however, as shown in
\figref{fig:N1_threshold} for the CLM. 
Prior to shear thickening, the average value of $N_1$ is nearly zero 
(slightly negative for CLM, slightly positive for ERM). 
The shear thickening transition is marked by two different behaviors,
depending on the volume fraction: 
at the lowest volume fractions studied, $N_1$  decreases across shear thickening,
while at larger $\phi$ there is a clear upturn towards positive
values at the shear thickening transition.
For the CLM, the behavior can be systematized by plotting $N_1$ as a
function of the stress, as in~\figref{fig:N1_threshold}: $N_1/\sigma$ is
an increasing function of $\sigma$, negative at low stresses, with a
qualitative behavior roughly independent of $\phi$.
The same plot for the ERM is similar but less systematic.
In any case, even above shear thickening, the ratio $N_1/\sigma$ is
always small, never exceeding $0.1$.


Overall, the behavior we observe for $N_1$ is in reasonable agreement
with most experiments~%
\citep{Lootens_2005, Lee_2006a, Larsen_2010, Couturier_2011, DBouk_2013}.
\citet{Zarraga_2000, Singh_2003, Dai_2013} report a negative value for $N_1$.
\citet{Lee_2006a} agree with our trend at low shear rates but observe
a subsequent change towards negative $N_1$ at very high shear rates,
while \citet{DBouk_2013} find a positive but significantly larger $N_1$.
In any case the amplitude of the fluctuations seems to be the relevant physical information
about $N_1$, as the average, positive or negative, is buried in the fluctuations in the time series,
such that at any time $N_1$ can be either positive or negative, 
even at large $\phi$ and $\dot \gamma$.

\subsection{Microstructure}

Non-Newtonian behavior of Brownian suspensions is often discussed
with shear induced microstructure; the particle configuration is
nearly at equilibrium at low P\'eclet number, while an anisotropic
microstructure is induced by shear at high P\'eclet number~\citep{Morris_2009}.
Ideas associated with the order-disorder transition scenario~%
\citep{Hoffman_1972, Hoffman_1974, Hoffman_1998} 
also predicted a clear structural change with shear thickening.
Experimental observations, however, revealed that the low viscosity state 
was not always ordered and that the shear thickening can occur with 
only a subtle signature in the microstructure~%
\citep{Bender_1996, Watanabe_1997, Watanabe_1998, Newstein_1999, Maranzano_2002}. 
One suggestion comes from the hydroclustering scenario,
which assumes the existence of hydrodynamically created extended density fluctuations.
Some experimental data are indeed in qualitative agreement 
with this scenario~\citep{Maranzano_2002,Cheng_2011}.


In the scenario we suggest in this article, 
the main structural modifications across the shear thickening transition 
should be sought in the contacts between particles. 
At low shear rates, 
frictional contacts are avoided because the particle pressure 
is too small to overcome the threshold (for the CLM) 
or the repulsion between particles (for the ERM). 
Those changes should thus be detected by measures specifically sensitive to the contacts. 
The pair correlation function 
should show few dramatic modifications across shear thickening.
This is what we show in this section.


We define the pair correlation function:
\begin{equation}
  \label{eq:pair_correl_defs}
  g_{\mathrm{all}}(\bm{r}) \equiv \frac{V}{N^2} \sum_{i,j} \delta(\bm{r} - \bm{r}_{ij}).
\end{equation}
The system being bidisperse, we can also define partial pair
correlation functions restricted to, 
e.g., pairs of small-small particles 
$g_{\mathrm{SS}}(\bm{r}) \equiv (V/N_{\mathrm{S}}^2)
\sum_{i,j\in \mathcal{S}_{\mathrm{S}}} \delta(\bm{r} - \bm{r}_{ij})$
where $\mathcal{S}_{\mathrm{S}}$ is the subset of $N_{\mathrm{S}}$ small particles. 
In the same way, we define the structure factor
(which is related to the pair correlation function via a Fourier Transform) as
\begin{equation}
  \label{eq:structure_factor_defs}
  S_{\mathrm{all}}(\bm{k}) 
  \equiv \frac{1}{N} \sum_{i,j\in \mathcal{S}, i\neq j}
  e^{i\bm{k}.\bm{r}_{ij}}.
\end{equation}

\begin{figure}[htb]
\centering
\includegraphics[width=0.9\textwidth]{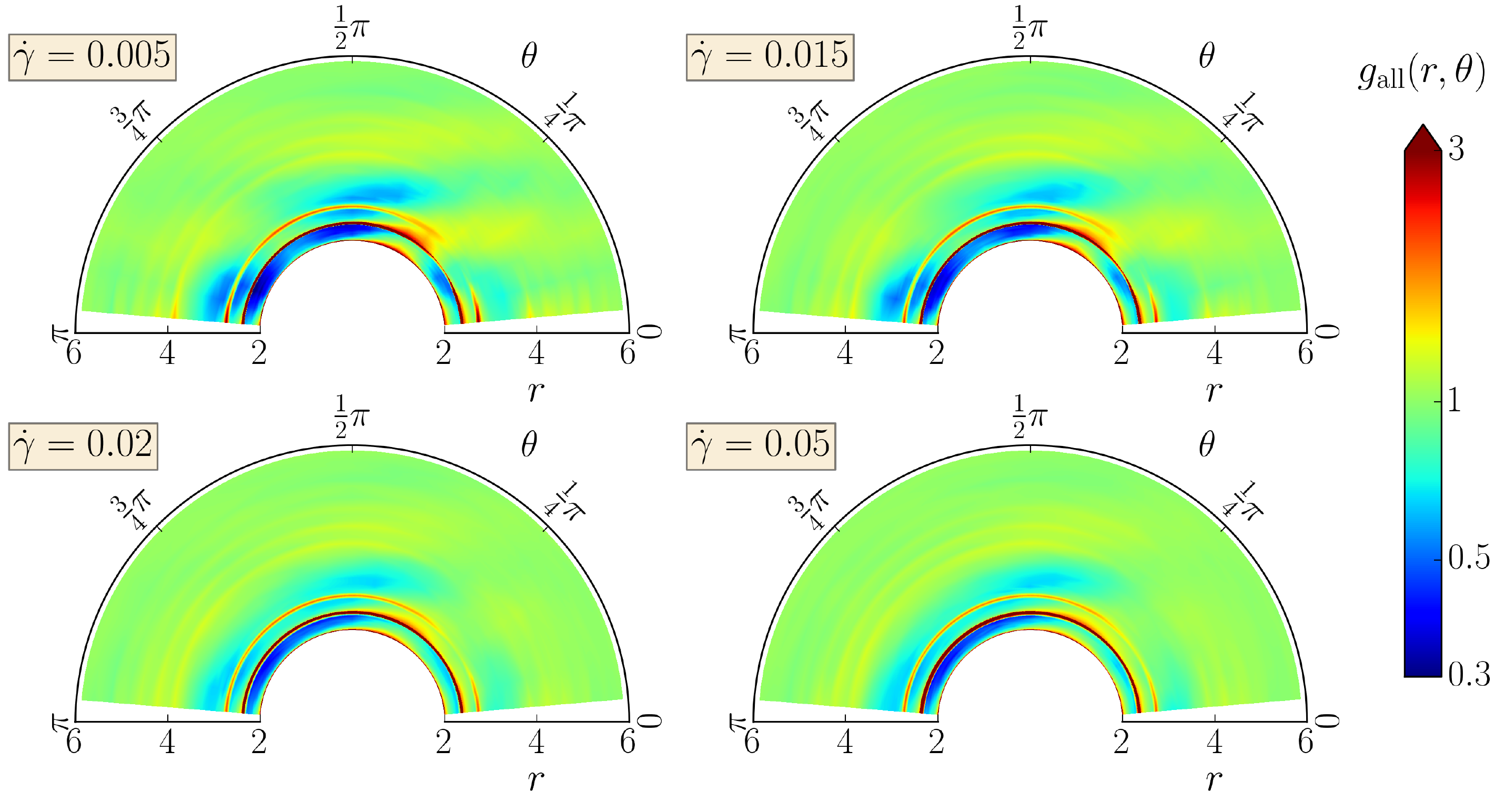} 
\caption{Pair correlation functions restricted in the shear plane
  $g_{\mathrm{all}}(r, \theta)$ for the CLM at $\phi=0.56$.
  The four plots correspond to shear rates $\dot \gamma = 0.005$ (well below DST),
  $\dot \gamma = 0.015$ (just below DST), $\dot \gamma = 0.02$ (just above DST), 
  and $\dot \gamma = 0.05$ (well above DST).}
\label{fig:gr_all_critical}
\end{figure}
\begin{figure}[htb]
\centering
\includegraphics[width=0.9\textwidth]{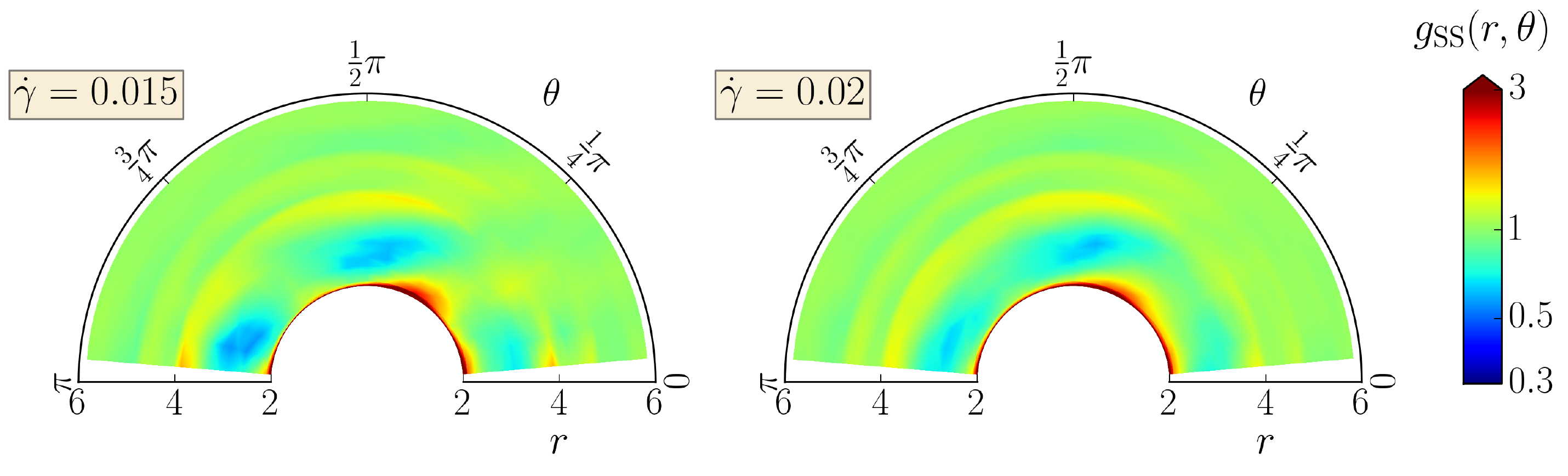} 
\caption{Pair correlation functions restricted to pairs of small
  particles in the shear plane $g_{\mathrm{SS}}(r, \theta)$ for the CLM at $\phi=0.56$. 
  The two plots correspond to shear rates 
  $\dot \gamma = 0.015$ (just below DST)
  and
  $\dot \gamma = 0.02$ (just above DST).
}
\label{fig:gr_ss_critical}
\end{figure}

In \figref{fig:gr_all_critical} and \figref{fig:gr_ss_critical}
we show the pair correlation function restricted 
to the shear plane (velocity/gradient) at $\phi=0.57$ 
for four values of the shear rate:
$\dot \gamma/\dot \gamma_0 = 0.005$, well below DST; 
$\dot \gamma/\dot \gamma_0 = 0.015$, just below DST; 
$\dot \gamma/\dot \gamma_0 = 0.02$, just above DST 
and $\dot \gamma/\dot \gamma_0 = 0.05$, well above DST.
As expected, the evolution of those plots seems weak around DST.
The main feature is a loss of contrast, 
unveiling a more isotropic structure above shear thickening.


\begin{figure}[htb]
\centering
\includegraphics[width=0.9\textwidth]{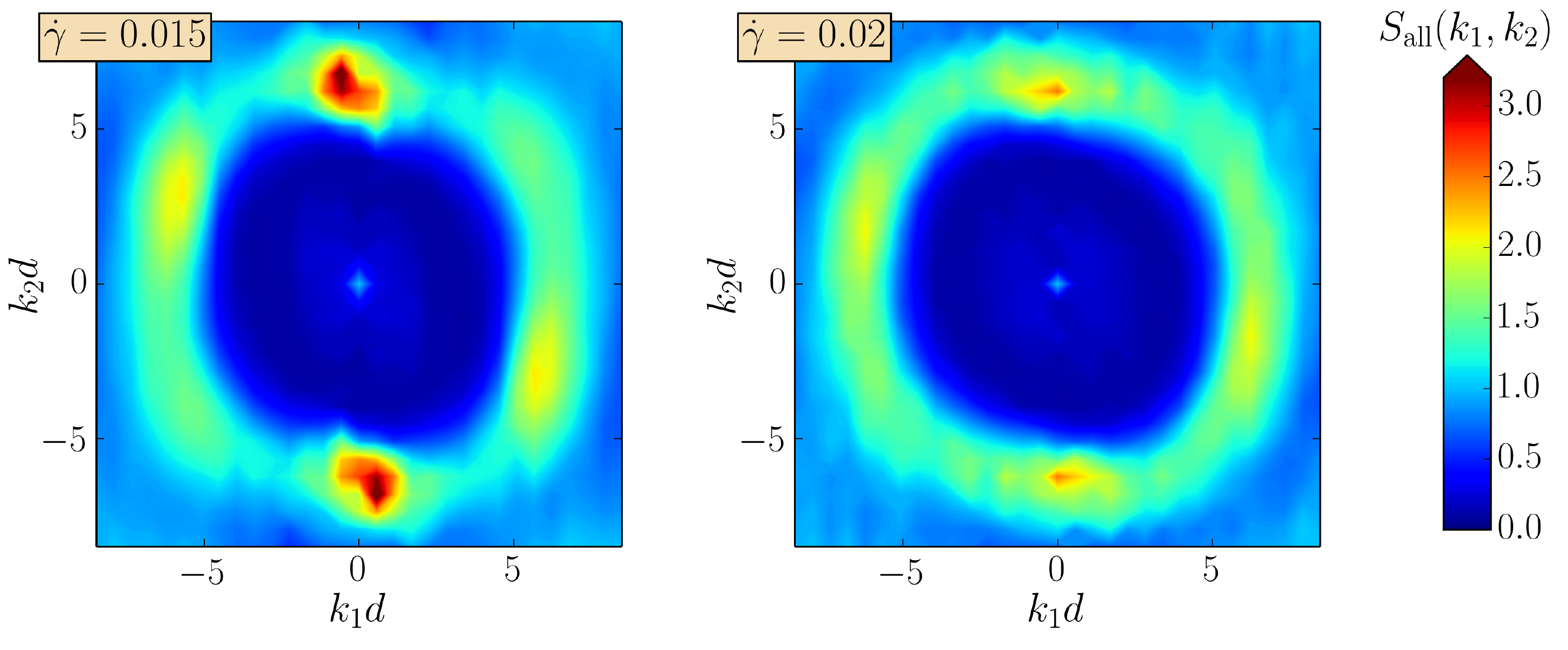} 
\caption{
  Structure factor in the shear plane $S_{\mathrm{all}}(k_1, k_2)$ 
  for the CLM at $\phi=0.56$.
  Left: $\dot\gamma = 0.015$ (just below DST). 
  Right: $\dot\gamma = 0.02$ (just above DST).
  The low viscosity state shows a peak in the gradient direction 
  (around $k_1 \approx 2\pi/d$) associated with a small amount of layering,
  which is strongly attenuated in the high viscosity phase.}
\label{fig:skshear_critical}
\end{figure}
\begin{figure}[htb]
  \centering
  \includegraphics[width=0.9\textwidth]{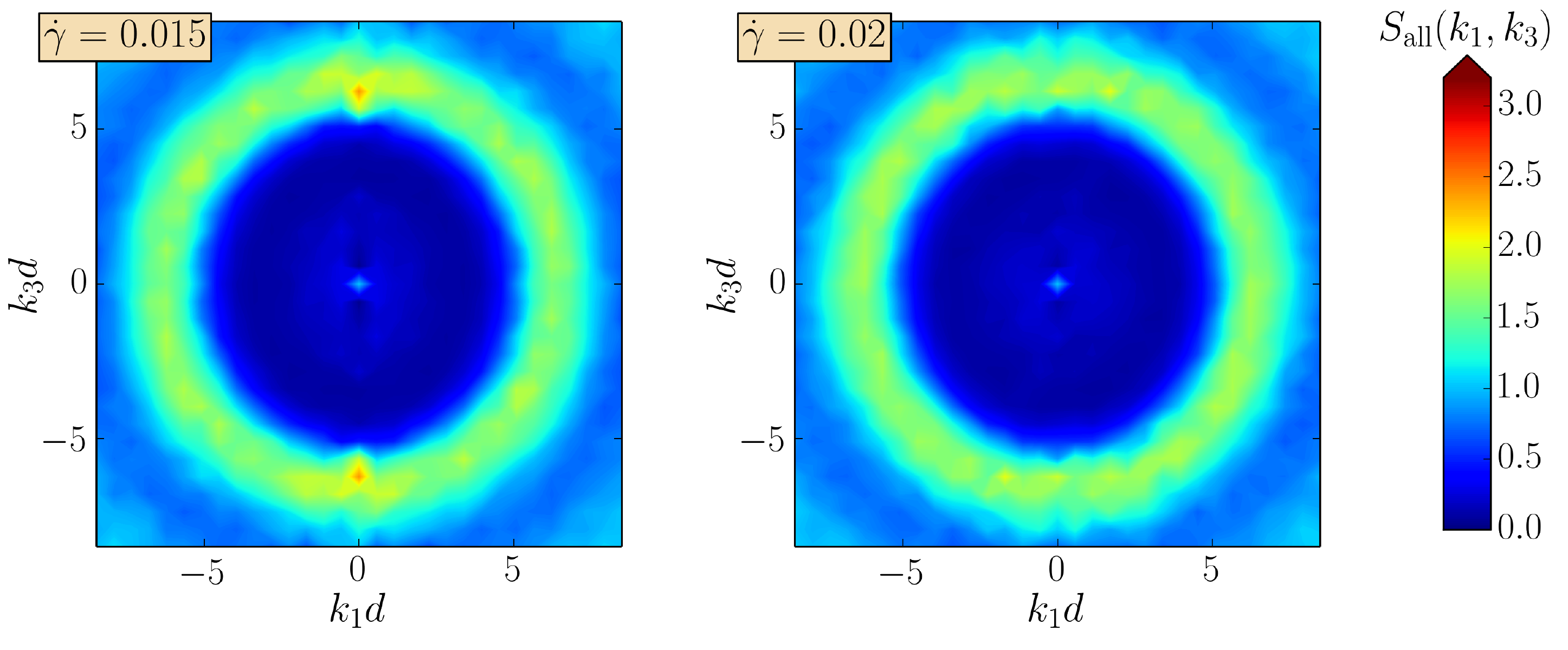} 
  \caption{Structure factor in the flow-vorticity plane 
    $S_{\mathrm{all}}(k_1, k_3)$ for the CLM at $\phi=0.56$.
    Left: $\dot \gamma = 0.015$ (just below DST).
    Right: $\dot \gamma = 0.02$ (just above DST).
    In this plane, the structure is isotropic in both states.
  }
  \label{fig:skfv_critical}
\end{figure}

The structure factor shown in~\figref{fig:skshear_critical} reveals another interesting feature: 
there is a clear anisotropy in the shear plane in the low viscosity state, 
with a peak in the gradient direction that is strongly reduced 
(but not absent, see the right of~\figref{fig:skshear_critical})
in the high viscosity state, above DST. 
This anisotropy is absent in the flow-vorticity plane; see~\figref{fig:skfv_critical}. 
Looking back at the pair correlation functions in~\figref{fig:gr_all_critical}, 
we see that the peak of $S_{\mathrm{all}}(\bm{k})$ along the gradient direction is the
signature of a slight ordering of the low viscosity phase:
there are small but visible downstream ripples aligned with 
the flow direction at small shear rate.

The slight layering we observe is however quite far from the
string order usually associated with the order-disorder scenario.
In the Supplementary Material, we show movies of the system projected
on the gradient-vorticity plane 
(i.e., looking down the stream direction)
for shear rates just below and just above shear thickening,
where we reduced the size of the particles to a third of their actual radius 
to allow a better visualization.
Those movies show warped layers along the flow-vorticity plane in some
parts of the system, while in other parts ordered layers are completely absent.
In any case, there is no further string ordering within the layers.
This is consistent with our structure factor data 
in the gradient-vorticity direction (not shown),
which exhibits only some structure in the gradient direction 
and no six-fold symmetry patterns typically observed 
when string ordering takes place~\citep{Laun_1992, Kulkarni_2009}.

\section{Discussion}

\subsection{Contact network}
\label{sec:contact_network}
\begin{figure}[htbp]
  \centering
  \includegraphics[width=14cm]{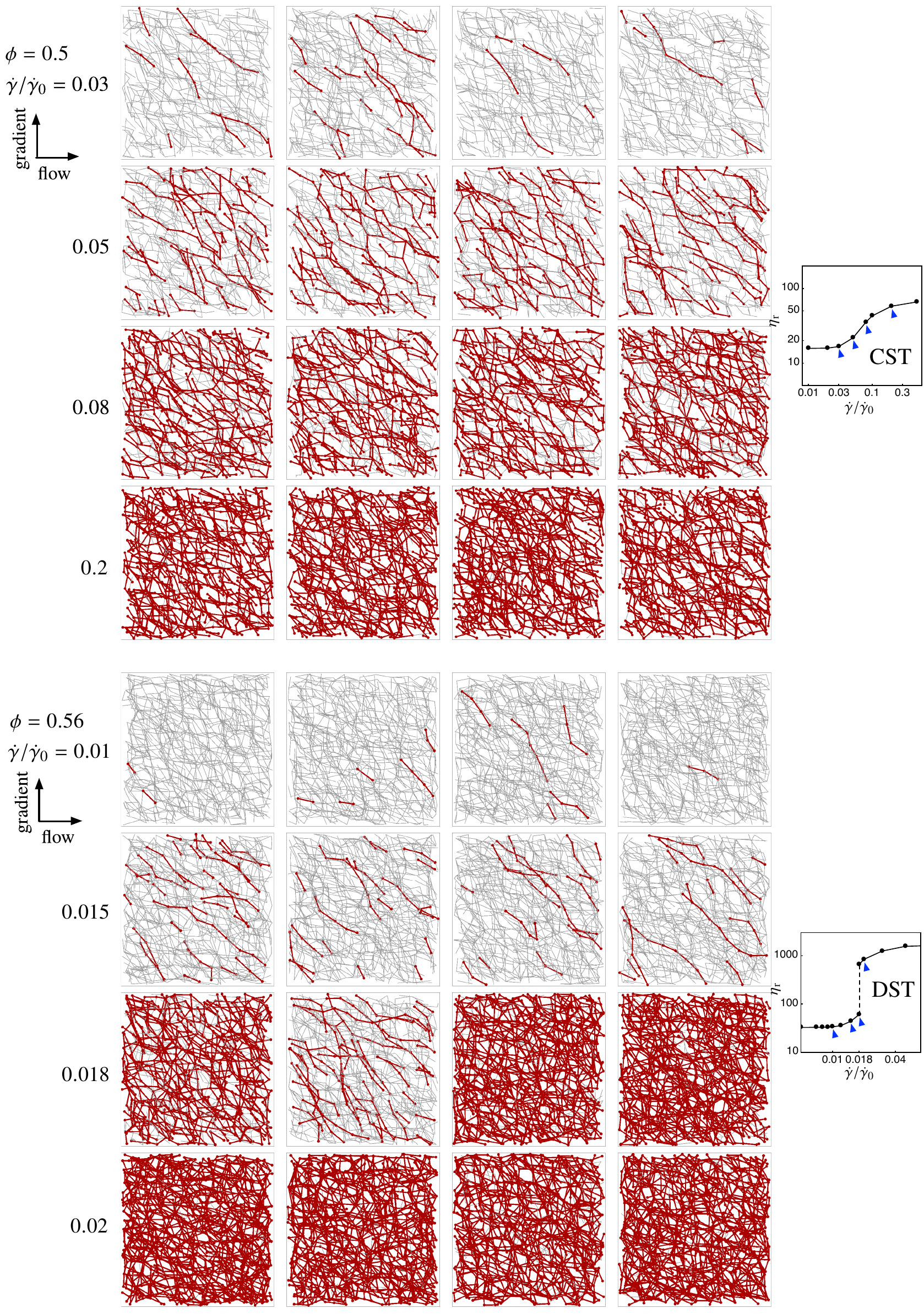}
  \caption{
    Snapshots of the contact network for the CLM at $\phi=0.5$ (top)
    and $0.56$ (bottom).
    Frictionless contacts (with normal force below the threshold
    $F_{\mathrm{CL}}$ are drawn in gray segments joining the centers
    of the two involved particles, while frictional contacts are drawn in red.
    Each row corresponds to a single
    shear rate, ranging from the low viscosity state to the high
    viscosity state and across CST (top) and DST (bottom).
    For each shear rate, we show four snapshots at
    different times (or equivalently strains) along the same
    simulation.
}\label{fig:contact_snapshots}
\end{figure}

In our simulation, 
the shear thickening transition is due to the appearance of a growing number 
of frictional contacts as the shear rate (or more precisely the shear stress) increases. 
The contact network that is built across shear thickening 
is shown in~\figref{fig:contact_snapshots}
for the CLM for the two cases of CST and DST, respectively.
(Two movies showing $\dot{\gamma}/\dot{\gamma}_0 = 0.015$ and $0.018$ of DST
are available in the Supplementary Material.
The time in these movies is scaled with $1/\dot{\gamma}$.)


At low shear rate, in the low viscosity state, 
frictional contacts only seldomly appear, 
being concentrated in small force chains along the compressional axis. 
At high shear rate, frictional contacts are the norm, 
creating a frictional contact network that is very close to being jammed;
i.e., having only a few (collective) degrees of freedom 
left to reorganize under the applied stress.


Between those two extreme situations, 
a whole continuum of gradually denser frictional contact networks 
is seen across the CST transition in~\figref{fig:contact_snapshots} (top). 
But in the case of DST in~\figref{fig:contact_snapshots} (bottom), 
the contact network discontinuously changes from sparse to dense at the transition, 
never showing configurations of intermediate densities.
Another interesting point is the occurrence of intermittency in the contact network, 
which is the immediate structural origin of the intermittent stress behavior shown
in~\figref{fig:stress_histograms_repulsion}: 
close to DST, the contact network suddenly switches from mostly frictionless to mostly frictional, 
showing a large sensitivity to fluctuations. 
When this happens, the viscosity immediately follows
by switching to high/low viscosity when the contact network respectively densifies/loosens.
\begin{figure}[htb]
  \includegraphics[width=9cm]{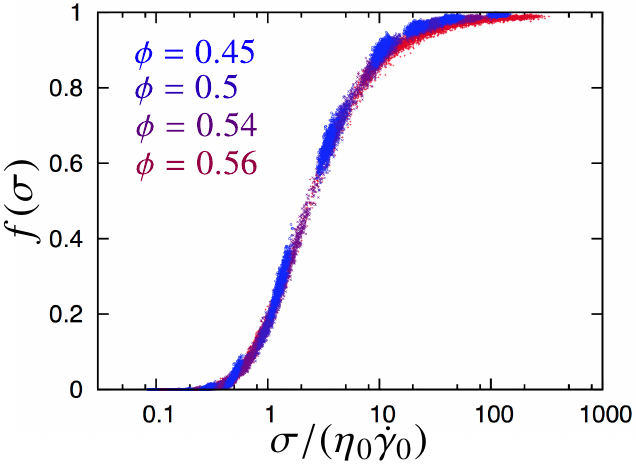}
  \caption{
    The fraction of frictional contacts $f$ as a function 
    of the shear stress $\sigma$, for several volume fractions. 
    There is a direct relation between 
    the stress and the proportion of frictional contacts,
    which emphasizes the fundamental role of friction 
    in shear thickening.
  }
  \label{fig:f_of_sigma}
\end{figure}


We can make the link between frictional contacts and shear stress
explicit by looking at the fraction of frictional contacts $f$~\citep{Wyart_2014}.
This quantity is unambiguously defined in the CLM model
as the ratio of the number of contacts $N_{\mathrm{C}}^{\mathrm{f}}$ 
that are in the frictional state 
(those for which the normal force exceeds the critical load $F_{\mathrm{CL}}$)
divided by the total number of contacts $N_{\mathrm{C}}$. 
This ratio has a direct relation to the shear stress, 
as the function $f(\sigma)$ plotted in~\figref{fig:f_of_sigma} shows.
This single relation demonstrates that, more than structural aspects, 
the shear thickening is above all 
a manifestation of the proliferation of frictional contacts in the system. 
We note one remarkable aspect of the relation $f(\sigma)$: 
it is independent of the volume fraction, 
at least in the range of $\phi$ that we have studied 
(which is rather large, owing to the fact that 
it should be understood as a range of distance to jamming $\phi-\phiFricJam$ 
rather than a bare range of $\phi$).
This peculiar aspect will be studied in a separate article.

\subsection{Phase diagram, relation to jamming}

While the jamming transition is 
the critical phenomenon underlying DST,
it should be noted that the jamming transition and DST are distinct,
and in particular $\phiDST \neq \phiFricJam$.
According to a recent theoretical argument by~\citet{Wyart_2014}, 
a scenario based on two diverging rheologies 
like the one we present in this work implies under reasonable assumptions
that $\phiDST < \phiFricJam$;
i.e., DST could then occur between two flowable (unjammed) states 
for volume fractions $\phiDST < \phi < \phiFricJam$.


In our simulations, 
we estimate $\phi_{\mathrm{J}}^{\mu=1} \approx 0.58$ 
(see~\figref{fig:two_rheologies}), 
while we observe DST for $\phi = 0.56$ 
(see~\figref{fig:visc_threshold}), 
which indeed implies $\phiDST < \phi_{\mathrm{J}}^{\mu=1}$.
For a volume fraction $\phi> \phi_{\mathrm{J}}^{\mu=1}$ 
only the low viscosity state would be visible under shear,
because the high viscosity frictional system is jammed at this volume fraction. 
This sets a maximum shear rate for the shear of the system at this volume fraction.


Physically, the two qualitatively different behaviors CST and DST, and
the fact that $\phiDST < \phi_{\mathrm{J}}^{\mu=1}$ can be explained
in our model as stemming from the level of contrast that exists
between frictionless and frictional rheologies, as suggested by~\citet{Wyart_2014} (part
of this explanation, when the contrast is diverging, was also
suggested by~\citet{Nakanishi_2012}).
Recalling that the fraction of frictional contacts $f(\sigma)$  essentially 
depends only on the applied shear stress (see the previous~\secref{sec:contact_network}),
and noting that the viscosity interpolates between 
the two rheologies according to the number of frictional contacts, 
we see that we can write a direct relation $\eta(\sigma)$ 
between the viscosity and the applied stress.
If the difference between the two rheologies
is small\textemdash i.e., $\eta(\sigma\to \infty)-\eta(\sigma\to 0)$ is
small\textemdash the curve $\eta(\sigma)$ is a sufficiently mildly increasing function 
such that $\dot\gamma=\sigma/\eta(\sigma)$ is also an increasing function of $\sigma$;
this corresponds to a single-valued curve $\sigma(\dot\gamma)$, and thus to CST. 
But if the difference $\eta(\sigma\to \infty)-\eta(\sigma\to 0)$ becomes too large,
$\eta(\sigma)$ increases faster than $\sigma$ in some interval, 
which means that $\dot\gamma(\sigma)$ is a decreasing function in this same interval.
This corresponds to a multivalued curve $\sigma(\dot\gamma)$, 
which is unstable, 
and shows up experimentally as a discontinuous curve, i.e., DST.


An increasing difference $\eta(\sigma\to \infty)-\eta(\sigma\to 0)$
is naturally provided in our model by the fact that the frictional
rheology diverges at a jamming volume fraction $\phiFricJam$
that is smaller than the frictionless rheology jamming point $\phiJam$. 
Then, at low volume fraction, the difference is small, 
but there must be a point $\phiDST$ below $\phiFricJam$ 
where the difference becomes large enough to observe a DST.

\begin{figure}[htb]
  \centering
\includegraphics[width=9cm]{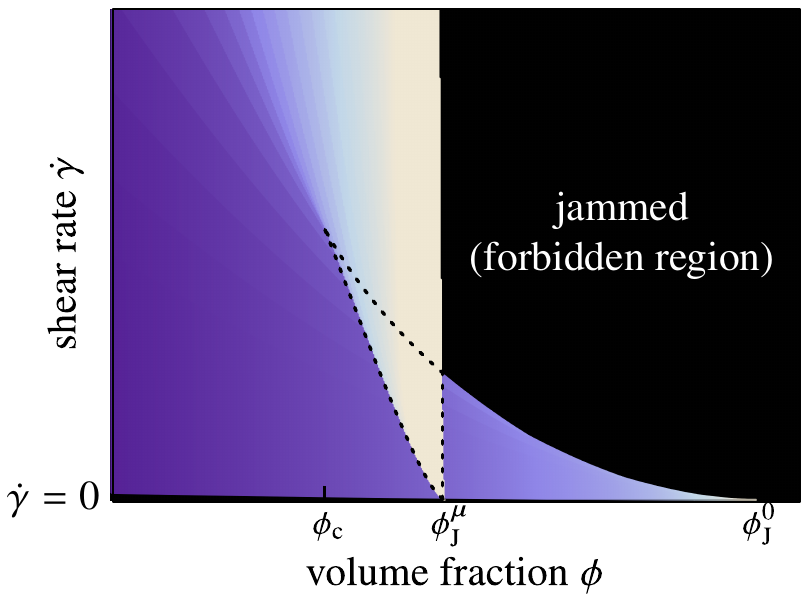}
\caption{
  Schematic phase diagram in the shear rate-volume fraction plane.
  The viscosity is color coded, from small (dark blue) to large values (white).
  Along the $\phi$ axis, 
  $\phi_{\mathrm{c}}$ is the point above which we can observe DST, 
  $\phiFricJam$ is the jamming point 
  for frictional spheres with friction coefficient $\mu$,
  and $\phiJam$ is the jamming point for frictionless spheres. 
  Intermittency is observed in
  the region delimited by a dashed line.}\label{fig:phase_diagram}
\end{figure}

These ideas can be summed up in
a phase diagram.
The rheology is essentially frictionless
in the lower part of the diagram~\figref{fig:phase_diagram},
as the stress is too small to activate friction between grains.
This rheology diverges 
at the frictionless jamming point $\phiJam$. 
The rheology is frictional in the upper part of the diagram, 
as friction is activated under the applied stress.
This rheology diverges at the frictional jamming point $\phiFricJam$ 
and thus shows a larger viscosity than the frictionless rheology.  
Thus two rheologies coexist 
on the shear rate-volume fraction plane.
Those rheologies are separated by a shear thickening 
that is continuous for $\phi<\phiDST$ 
and discontinuous for $\phiDST< \phi< \phiFricJam$.


The discontinuity is actually related to the coexistence of 
the two rheologies in the triangle delimited by a dashed line.
In this region, we observe one consequence of the coexistence in the
intermittency of the flow: 
the system switches from one state to the other through activation events, 
as is shown by the time series of the stress 
(see~\figref{fig:stress_histograms_repulsion}).  
For $\phiFricJam < \phi < \phiJam$,
DST is strictly speaking no longer observed, 
as the high viscosity flowable state does not exist any more.
In this region, 
DST is actually replaced by shear jamming~\citep{Bi_2011}:
if one applies a shear stress larger than $\sigma_{\mathrm{ST}}$, 
the system goes to a solid state and cannot flow.
This implies the existence of a forbidden region, in gray,
where no flow is possible for hard spheres.
In the lower shear-rate domain above $\phiFricJam$,
the low viscosity state is stable, but the activated events responsible
for the intermittency at lower $\phi$ in small
systems might lead to complete jamming in a finite time.
Lastly, the dot in the upper left corner of the DST region is a
critical point separating CST from DST.
It shares some features with a critical point of a second order phase
transition, for instance a diverging susceptibility
$\mathrm{d}\sigma/\mathrm{d}\dot\gamma$.


Of course, as we numerically work with a shear-rate controlled scheme,
the system always flows at any shear rate and any volume fraction,
even in the forbidden region of the phase diagram.
But in the latter region flow is only achieved by creating large overlaps
between the particles (i.e., compressing them), hence violating the
criteria we set to mimic hard sphere suspensions.
At high shear rate and for $\phi > \phiFricJam$, 
a real hard sphere system would not flow,
but it does in the simulation because of our inability to enforce the
hard sphere condition in a high (possibly infinite) stress state. 
The spheres we simulate in that case cannot be considered hard any more,
and their stiffness sets a cutoff to the stress scales under shear. 
This situation is somehow similar to the one observed in experiments,
where it is observed that the stress in the high viscosity state 
is set by the weakest stress scale in the sample 
(which might be a large stress scale for the rheometer usage range),
whether it is the particles' stiffness or the surface tension 
on the edge of the rheometer~\citep{Brown_2012}.
Therefore, in an experiment, 
even if the idealized system would be jammed under the investigated conditions, 
the real system still flows, perhaps thanks to dilation, whereas in our simulation, 
under the same conditions, it still flows thanks to the deformability of particles.


The difference between $\phiDST$ and $\phiFricJam$ may have been observed 
in some stress drop experiments~\citep{OBrien_2000, Larsen_2010}. 
For $\phiDST < \phi < \phiFricJam$, 
the phase diagram of~\figref{fig:phase_diagram} predicts 
that the stress in the thickened state 
would relax quickly upon flow cessation,
as the contact network is unjammed.
By contrast, for $\phiFricJam < \phi < \phiJam$
the stress would not relax entirely, 
as part of it would be stored elastically in the jammed contact network.
This scenario is actually consistent with the data
from~\citep{OBrien_2000, Larsen_2010}.

\section{Conclusion}
\label{sec:conclusion}

We have introduced a frictional-viscous model of dense suspensions under shear
that is built on the framework used by previous standard models of suspensions: 
simulations are performed in the zero particle-inertia
and zero fluid-inertia limits ($\StNum \to 0$ and $\ReNum \to 0$), 
and include relevant hydrodynamic interactions
and a short-range repulsive potential.
The critical innovation is that, besides this purely hydrodynamic basis, 
it includes frictional contacts between solid particles.
This model can be used to simulate sheared suspensions over a range of volume fractions,
from mildly dense, where the short range hydrodynamic interactions dominate,
to the jamming point, where contact forces dominate.


The effect of adding frictional contacts is striking at large volume fractions: 
tangential forces due to friction restrict microscopic rearrangements in the system,
resulting in a large shear stress. 
The number of contacts created during the flow is 
the result of a competition between applied shear forces 
and the short-range repulsive forces.
Therefore, alongside the contactless (hence frictionless) 
rheology at low shear rates
a new contact dominated frictional
rheology appears at high shear rates.
Shear thickening is then simply the transition from the contactless
rheology to the contact dominated rheology with increasing shear
stress (and shear rate).
The strongest evidence that this stress-induced friction scenario is correct
is the ability of our model to reproduce shear thickening, 
including its discontinuous form, and its volume-fraction dependence.
Therefore, we conclude that friction is a key element for shear thickening.


The shear thickening mechanism we describe is qualitatively
independent of our parameter choices.
Quantitatively, we can summarize the effects of the various parameters as follows.
The friction coefficient $\mu$ only affects the volume fraction range over
which the phenomenon happens:
if $\mu$ increases, the critical volume fraction $\phi_{\mathrm{c}}$ decreases.
Any friction coefficient $\mu>0$ leads to a DST,
with $\phi_{\mathrm{c}}(\mu)\to \phi_{\mathrm{J}}^{0}$ in the limit $\mu\to 0$.
The lubrication cutoff $\delta$ mainly sets the dissipation level
in the systems and thus sets an overall prefactor in the entire curve $\eta(\sigma)$:
the viscosity is larger when $\delta$ decreases.
For instance, we performed simulations with a cutoff $\delta=10^{-2}$
and compared the results to the ones shown in the present work obtained
with $\delta=10^{-3}$: the curve $\eta^{\delta=10^{-2}}(\sigma)$ can
be superimposed to $\eta^{\delta=10^{-3}}(\sigma)$ by a mere
multiplicative factor
$\eta^{\delta=10^{-2}}(\sigma) \approx 0.8\eta^{\delta=10^{-3}}(\sigma)$.
This has the effect of pushing the shear thickening to lower shear rates,
as the onset stress $\sigma_{\mathrm{ST}}$ is independent of $\delta$.
The particle stiffness constants $k_{\mathrm{n}}$ and $k_{\mathrm{t}}$
are chosen such that the simulations are run in the asymptotic regime
where the dimensionless numbers $6 \pi \eta_0 a\dot{\gamma}/k_{\mathrm{n, t}}$
are very small.
In this asymptotic regime,
the rheology is independent of these stiffness constants.
The Debye length $\kappa^{-1}$ governs the shear thinning behavior
at low shear rates.
Our Critical Load Model can be thought as having a vanishing Debye length.
Comparing with the Electrostatic Repulsion Model with $\kappa^{-1} = 0.05a$
(with $a$ the radius of the small particles),
we observe that besides the shear thinning regime,
a finite $\kappa^{-1}$ virtually does not affect the shear thickening and
the high viscosity state.


We now turn to a quantitative comparison with experimental data.
The main tuning parameter in our description is the force scale
$F_{\mathrm{CL}}$ for CLM (or $F_{\mathrm{ER}}$ for the ERM).
This parameter sets the onset stress at which shear thickening occurs.
We can estimate this force for a system of $\SI{1}{\micro\meter}$ silica spheres
in water.
We find an electrostatic repulsion at contact of order
$F_{\mathrm{ER}}\approx \SI{0.2}{\nano\newton}$
(as evaluated in~\citet{Behrens_2001} with a charge density
$\approx \SI{5d-4}{\coulomb\per\square\meter}$
and a Debye length $\kappa^{-1} \approx \SI{0.1}{\micro\meter}$).
This gives for our model the stress scale $\eta_0 \dot{\gamma}_0 =
(6\pi a^2)^{-1} F_{\mathrm{ER}}$, which together with the
non-dimensionalized onset stress $\sigma_{\mathrm{ST}}/(\eta_0
\dot{\gamma}_0) \approx 0.3$ obtained the ERM model
(see~\figref{fig:visc_repulsion}) gives a predicted onset stress of
$\sigma_{\mathrm{ST}} \approx \SI{8}{\pascal}$.  
For the very same system, \citet{Lootens_2005} experimentally find
an onset stress of $\sigma_{\mathrm{ST}} \approx \SI{5}{\pascal}$,
with which our simulation is thus in good agreement.
Thus, our mechanism seems to give a \emph{quantitative}
description of shear thickening in dense suspensions,
even if it would require comparison to more experimental data
to be definitely conclusive.
(Obtaining a reliable value of the microscopic force scale
  is rarely feasible in the published data on DST,
  and this restricts the number of comparisons we can make at the moment.)


Although our model assumes non-Brownian suspensions,
we may expect that the same mechanism explains shear thickening
of Brownian colloidal suspensions.
Indeed, the Brownian forces qualitatively play a role similar to the
repulsive potential that we use in our ERM:
at low shear stress, the Brownian forces are large enough to randomly open gaps
between contacting particles, while at high shear stress this becomes much less likely,
as relative trajectories of pairs of particles become smooth
at high Peclet numbers~\citep{Nazockdast_2013}.
The fact that a random noise can relax friction between grains is a
known phenomenon in granular matter~\citep{Gao_2009,Karim_2014}.
Friction would then develop in a Brownian suspension in a manner very
similar to the present work.

\section*{Softwares and scripts}
The code used to generate the data is available at
\url{https://bitbucket.org/rmari/lf_dem}. 
Our code makes use of the
CHOLMOD library v2.0.1 by Tim Davis
(\url{https://www.cise.ufl.edu/research/sparse/cholmod/}) for direct
Cholesky factorization of the sparse resistance matrix.
Part of the data analysis and figure plotting was done using \mbox{numpy}
v1.8.0, \mbox{matplotlib} v1.3.1 and \mbox{IPython} v1.1.0.
The associated data and the IPython notebook are available at
\url{https://github.com/rmari/Data_Paper_JOR_58_1693_2014}.
The notebook can be visualized without IPython at
\url{http://nbviewer.ipython.org/github/rmari/Data_Paper_JOR_58_1693_2014/blob/master/JOR_figures_Rheology_Data.ipynb}.

\section*{ACKNOWLEDGMENTS}

We would like to thank Mike Cates and Matthieu Wyart for the many
discussions concerning our respective work and Philippe Claudin for
discussions about the $\mu(I_{\mathrm{v}})$ rheology.
This research was supported in part by a grant of computer time from the City
University of New York High Performance Computing Center under NSF
Grants CNS-0855217, CNS-0958379, and ACI-1126113.
J. F. M. was supported in part by NSF PREM (DMR 0934206).

\appendix

\section{Hydrodynamic resistances}
\label{app:resistance_matrices}

The hydrodynamic interactions
arising from an imposed background flow and relative motions 
between nearby particles are given by
\begin{equation}
  \begin{pmatrix}
    \bm{F}_{\mathrm{H}} \\
    \bm{T}_{\mathrm{H}}
  \end{pmatrix}
  =
  - (\bm{R}_{\mathrm{Stokes}}+\bm{R}_{\mathrm{Lub}})\cdot
  \begin{pmatrix}
    \bm{U}-\bm{U}^{\infty}\\
    \bm{\Omega}-\bm{\Omega}^{\infty}\\
  \end{pmatrix}
  + 
  \bm{R}'_{\mathrm{Lub}}:
  \bm{E}^{\infty},
\end{equation}
where a diagonal matrix $\bm{R}_{\mathrm{Stokes}}$ 
comes from the (one-body) Stokes drag 
and sparse matrices $\bm{R}_{\mathrm{Lub}}$ and $\bm{R}'_{\mathrm{Lub}}$
come from the (two-body) lubrication. 


Using the basic units $L_0 \equiv a$ for lengths, $U_0 \equiv L_0 \dot{\gamma}$ for velocities,
and $F_0 \equiv 6 \pi \eta_0 L_0 U_0 $ for forces, 
the elements of $\bm{R}_{\mathrm{Stokes}}$ give the Stokes drag through
\begin{equation}
    \begin{pmatrix}
      \bm{F}_{\mathrm{Stokes}}^{(i)}\\
      \bm{T}_{\mathrm{Stokes}}^{(i)}      
    \end{pmatrix}
    = - \bm{R}^{(i,i)}_{\mathrm{Stokes}} \cdot
    \begin{pmatrix}
    \bm{U}^{(i)}-\bm{U}^{\infty}(\bm{r}^{(i)})\\
    \bm{\Omega}^{(i)}-\bm{\Omega}^{\infty}
    \end{pmatrix}
    = -
    \begin{pmatrix}
    \bm{U}^{(i)}-\bm{U}^{\infty}(\bm{r}^{(i)})\\
    (4/3)(\bm{\Omega}^{(i)}-\bm{\Omega}^{\infty})
    \end{pmatrix}
    ,
\end{equation}
while $\bm{R}_{\mathrm{Lub}}$ and $\bm{R}'_{\mathrm{Lub}}$ consist of
off-diagonal blocks giving the lubrication forces and torques
for a pair $(i,j)$ through
\begin{equation}
  \label{eq:explicit_resmat}
  \begin{pmatrix}
    \bm{F}_{\mathrm{H}}^{(i, j)}\\
    \bm{F}_{\mathrm{H}}^{(j, i)}\\
    \bm{T}_{\mathrm{H}}^{(i, j)}\\
    \bm{T}_{\mathrm{H}}^{(j, i)}
  \end{pmatrix}
  =
  - \bm{R}_{\mathrm{Lub}}^{(i, j)}\cdot
  \begin{pmatrix}
    \bm{U}^{(i)}-\bm{U}^{\infty}(\bm{r}^{(i)})\\
    \bm{U}^{(j)}-\bm{U}^{\infty}(\bm{r}^{(j)})\\
    \bm{\Omega}^{(i)}-\bm{\Omega}^{\infty}\\
    \bm{\Omega}^{(j)}-\bm{\Omega}^{\infty}
  \end{pmatrix}
  + \bm{R}_{\mathrm{H}}^{\prime(i, j)}:
  \begin{pmatrix}
    \bm{E}^{\infty}\\
    \bm{E}^{\infty}
  \end{pmatrix}
  .
\end{equation}


Following the notation of \citet{Jeffrey_1984,Jeffrey_1992}, 
the matrices $\bm{R}_{\mathrm{Lub}}^{(i, j)}$
and $\bm{R}_{\mathrm{Lub}}^{\prime(i, j)}$ are obtained 
from the particle positions as 
(we give only the upper triangular part of the symmetric $\bm{R}_{\mathrm{Lub}}^{(i, j)}$):
\begin{gather}
  \bm{R}_{\mathrm{Lub}}^{(i, j)}
  \equiv
  \begin{pmatrix}
    X^A_{ii} \operatorPn + Y^A_{ii} \operatorPt &
    X^A_{ij} \operatorPn + Y^A_{ij} \operatorPt &
    Y^B_{ii} \operatorPr & 
    Y^B_{ji} \operatorPr \\
    . &
    X^A_{jj} \operatorPn + Y^A_{jj} \operatorPt &
    Y^B_{ij} \operatorPr &
    Y^B_{jj} \operatorPr \\
    . &
    . &
    Y^C_{ii} \operatorPt &
    Y^C_{ij} \operatorPt \\
    . &
    . &
    . &
    Y^C_{jj} \operatorPt
\end{pmatrix}
, \\
\bm{R}_{\mathrm{Lub}}^{\prime(i, j)}
\equiv
\begin{pmatrix}
  X^G_{ii} \operatorname{Q}_{\bm{n}_{ij}} + Y^G_{ii} \operatorname{Q}_{\bm{n}_{ij}}' &
  X^G_{ji} \operatorname{Q}_{\bm{n}_{ij}} + Y^G_{ji} \operatorname{Q}_{\bm{n}_{ij}}' \\
  X^G_{ij} \operatorname{Q}_{\bm{n}_{ij}} + Y^G_{ij} \operatorname{Q}_{\bm{n}_{ij}}' & 
  X^G_{jj} \operatorname{Q}_{\bm{n}_{ij}} + Y^G_{jj} \operatorname{Q}_{\bm{n}_{ij}}' \\
  Y^H_{ii} \operatorname{Q}_{\bm{n}_{ij}}^{\mathrm{r}} &
  Y^H_{ji} \operatorname{Q}_{\bm{n}_{ij}}^{\mathrm{r}}  \\
  Y^H_{ij} \operatorname{Q}_{\bm{n}_{ij}}^{\mathrm{r}} & 
  Y^H_{jj} \operatorname{Q}_{\bm{n}_{ij}}^{\mathrm{r}}
\end{pmatrix}
.
\end{gather}
In these expressions, we have introduced the normal projection operator
$\operatorPn \equiv \bm{n}_{ij} \bm{n}_{ij}$,
the tangential projection operator $\operatorPt\equiv \id - \bm{n}_{ij} \bm{n}_{ij}$,
and the ``cross product'' operator $\operatorPr$ defined as 
$\operatorPr \bm{q} \equiv \bm{n}_{ij} \times \bm{q}$.
We also used the operators $\operatorname{Q}_{\bm{n}_{ij}}$,
$\operatorname{Q}_{\bm{n}_{ij}}'$ and
$\operatorname{Q}_{\bm{n}_{ij}}^{\mathrm{r}}$ defined for an
arbitrary matrix $\bm{M}$ as:
\begin{equation}
  \begin{split}
    \operatorname{Q}_{\bm{n}_{ij}} \bm{M}  
    & \equiv \biggl(\bm{M}:\bm{n}_{ij}\bm{n}_{ij} 
      - \frac{1}{3}\operatorname{Tr}\bm{M} \biggr) \bm{n}_{ij}, \\
    \operatorname{Q}_{\bm{n}_{ij}}' \bm{M}  
    & \equiv \bigl( \bm{M} + \bm{M}^{\mathrm{T}} \bigr) 
    \cdot \bm{n}_{ij} - 2(\bm{M}:\bm{n}_{ij}\bm{n}_{ij}) \bm{n}_{ij} , \\
    \operatorname{Q}_{\bm{n}_{ij}}^{\mathrm{r}} \bm{M}  
    & \equiv
    2 \bm{n}_{ij}\times \left[\bigl( \bm{M} + \bm{M}^{\mathrm{T}} \bigr) 
      \cdot \bm{n}_{ij}\right].
  \end{split}
\end{equation}
The scalar coefficients $X$ and $Y$ have an explicit dependence 
on the non-dimensional interparticle gap 
$h^{(i,j)} \equiv 2(r^{(i,j)} - a_i- a_j)/(a_i+a_j)$,
and we use only the terms of leading order:
\begin{equation}
  \label{eq:leading_terms}
  X \equiv g^{X}\frac{1}{h^{(i,j)}+\delta}, 
  \quad
  Y \equiv g^{Y}\log\frac{1}{h^{(i,j)}+\delta}.
\end{equation}
With $\lambda \equiv a_j/a_i$, 
the coefficients $g^{X}$ and $g^{Y}$ appearing 
in $\bm{R}_{\mathrm{Lub}}^{(i, j)}$ are written
\begin{equation}
  \label{eq:leading_terms_RFU}
  \begin{aligned}
  g^{X^A_{ii}}(\lambda) &= 2 a_i \frac{\lambda^2}{(1+\lambda)^3},
  & 
  g^{X^A_{jj}}(\lambda) &= \lambda g^{X^A_{ii}}(\lambda^{-1}), \\
  g^{X^A_{ij}}(\lambda) &= - 2 (a_i+a_j) \frac{\lambda^2}{(1+\lambda)^4},
  & 
  g^{X^A_{ji}}(\lambda) &= g^{X^A_{ij}}(\lambda^{-1}) = g^{X^A_{ij}}(\lambda),\\
  g^{Y^A_{ii}}(\lambda) &= \frac{4 a_i}{15} \frac{\lambda(2+\lambda +2\lambda^2)}{(1+\lambda)^3},
  & 
  g^{Y^A_{jj}}(\lambda) &= \lambda g^{Y^A_{ii}}(\lambda^{-1}),\\
  g^{Y^A_{ij}}(\lambda) &= - \frac{4 (a_i+a_j)}{15} 
  \frac{\lambda(2+\lambda +2\lambda^2)}{(1+\lambda)^4},
  & 
  g^{Y^A_{ji}}(\lambda) &= g^{Y^A_{ij}}(\lambda^{-1}) = g^{Y^A_{ij}}(\lambda),\\
  g^{Y^B_{ii}}(\lambda) &= -\frac{2 a_i^2}{15} \frac{\lambda(4+\lambda)}{(1+\lambda)^2},
  & 
  g^{Y^B_{jj}}(\lambda) &= - \lambda^2 g^{Y^B_{ii}}(\lambda^{-1}),\\
  g^{Y^B_{ij}}(\lambda) &= \frac{2 (a_i+a_j)^2}{15} \frac{\lambda(4+\lambda)}{(1+\lambda)^4},
  & 
  g^{Y^B_{ji}}(\lambda) &= - g^{Y^B_{ij}}(\lambda^{-1}),\\
  g^{Y^C_{ii}}(\lambda) &= \frac{8 a_i^3}{15} \frac{\lambda}{1+\lambda},
  & 
  g^{Y^C_{jj}}(\lambda) &= 
  \lambda^3 g^{Y^C_{ii}}(\lambda^{-1}) = \lambda^2 g^{Y^C_{ii}}(\lambda),\\
  g^{Y^C_{ij}}(\lambda) &= \frac{2 (a_i+a_j)^3}{15} \frac{\lambda^2}{(1+\lambda)^4},
  & 
  g^{Y^C_{ji}}(\lambda) &= g^{Y^C_{ij}}(\lambda^{-1}) = g^{Y^C_{ij}}(\lambda).
  \end{aligned}
\end{equation}
Similarly, the terms appearing in $\bm{R}_{\mathrm{Lub}}^{\prime(i, j)}$ are
\begin{equation}
  \label{eq:leading_terms_RFE}
  \begin{aligned}
    g^{X^G_{ii}}(\lambda) &= 2 a_i^2 \frac{\lambda^2}{(1+\lambda)^3},
    & 
    g^{X^G_{jj}}(\lambda) &= - \lambda^2 g^{X^G_{ii}}(\lambda^{-1}),\\
    g^{X^G_{ij}}(\lambda) &= - 2 (a_i+a_j)^2 \frac{\lambda^2}{(1+\lambda)^5},
    & 
    g^{X^G_{ji}}(\lambda) &= - g^{X^G_{ij}}(\lambda^{-1}),\\
    g^{Y^G_{ii}}(\lambda) &= \frac{a_i^2}{15} \frac{4\lambda -
      \lambda^2 +7\lambda^3}{(1+\lambda)^3},
    & 
    g^{Y^G_{jj}}(\lambda) &= - \lambda^2 g^{Y^G_{ii}}(\lambda^{-1}),\\
    g^{Y^G_{ij}}(\lambda) &= - \frac{(a_i+a_j)^2}{15} \frac{4\lambda -
      \lambda^2 +7\lambda^3}{(1+\lambda)^5},
    & 
    g^{Y^G_{ji}}(\lambda) &= - g^{Y^G_{ij}}(\lambda^{-1}),\\
    g^{Y^H_{ii}}(\lambda) &= \frac{2a_i^3}{15} \frac{2\lambda -
      \lambda^2}{(1+\lambda)^2},
    & 
    g^{Y^H_{jj}}(\lambda) &=  \lambda^3 g^{Y^G_{ii}}(\lambda^{-1}),\\
    g^{Y^H_{ij}}(\lambda) &= \frac{(a_i+a_j)^3}{15} \frac{\lambda^2 +7\lambda^3}{(1+\lambda)^5},
    & 
    g^{Y^H_{ji}}(\lambda) &=  g^{Y^H_{ij}}(\lambda^{-1}).
  \end{aligned}
\end{equation}

\section{Contact model}
\label{app:contact_model_full}

\subsection{Tuning the spring constants}

Our objective is to study the rheology of hard-sphere suspensions.
We have no way to mimic rigorously hard spheres
by using a contact model with springs.
To provide the best approximation to hard spheres, the elastic
constants appearing in the contact model must satisfy a simple constraint:
they should be large enough to generate as little geometric overlap as
possible between the particles.
(An equivalent way to state this is through the
requirement that the contact based non-dimensionalized shear rate 
$6 \pi \eta_0 a\dot{\gamma}/k_{\mathrm{n}}\ll 1$ 
introduced in~\secref{sec:shearrate_dep} is sufficiently small.)


We therefore set a criterion for the overlap: 
the largest overlap  between any two particles 
during the simulation should not be larger than $h_{\mathrm{max}}$ 
(\SI{5}{\percent} of the particle radius in this simulation).
As the overlap $|h|$ depends on the shear stress, 
an estimate being $\langle |h| \rangle \sim (1/k_{\mathrm{n}}) \sigma(\phi,\dot\gamma) a^2$,
the spring constant has to be tuned for every $\phi$ and $\dot\gamma$ 
so that $\max(|h|) < h_{\mathrm{max}}$.
%
We introduce a similar criterion for the tangential spring 
stretch mimicking static friction: 
we pick $ k_{\mathrm{t}}(\phi, \dot\gamma) $ 
so that $ \xi $ is smaller than $ \SI{5}{\percent} $ of the particle radius. 
We detail below how we fulfill these two conditions and retain hard-sphere 
like behavior in our simulation.


For a given $\phi$, the largest stress is obtained in the high shear-rate limit.
Hence, we start by determining high shear-rate spring constants 
$\bigl(k_{\mathrm{n}}^{\ast}(\phi),k_{\mathrm{t}}^{\ast}(\phi)\bigr)$ 
by running pre-simulations at $\dot\gamma\to \infty$, 
where these values are regularly updated with a certain interval 
until the criteria on the overlap and the tangential spring stretch are fulfilled.

Second, a trivial shear-rate dependence comes in if the contact model
introduces a force scale in addition to the hydrodynamic one.
Avoiding this requires picking the shear rate dependence of
$k_{\mathrm{n}}(\phi, \dot\gamma)$ and 
$k_{\mathrm{n}}(\phi, \dot\gamma)$ 
by scaling these parameters with the shear rate; 
i.e.,
$k_{\mathrm{n}}(\phi, \dot{\gamma}) = \dot{\gamma} k_{\mathrm{n}}^{\ast}(\phi) $
and 
$k_{\mathrm{t}}(\phi, \dot{\gamma}) = \dot{\gamma} k_{\mathrm{t}}^{\ast}(\phi) $.
With this scaling, there is no competition between hydrodynamic and contact forces,
as they are proportional, 
so an additional explicit force scale (as the one in~\secref{sec:repulsion}) 
is required to have a shear-rate dependence in our simulation.
Also note that as the high shear-rate limit is always 
the largest viscosity at a given $\phi$, 
with this choice of scaling 
$k_{\mathrm{n}}(\phi, \dot{\gamma})$ and $k_{\mathrm{t}}(\phi, \dot{\gamma})$ 
always fulfill the criteria we set for the overlap and tangential spring stretch.


\subsection{Combining the overdamped dynamics and the contact model}

In this appendix we present a slightly more general contact model
than the one presented in the main text. 
While the contact model may not deserve a lengthy description in itself,
immersing an orthodox contact model in overdamped dynamics leads to
non-trivial difficulties, 
which justifies our use of a simpler version of the model.


A more general stick/slide friction model using springs and dashpots
is the following~\citep{Cundall_1979, Luding_2008}:
\begin{equation}
  \begin{split}
    \bm{F}_{\mathrm{C,nor}}^{(i,j)} & = k_{\mathrm{n}} h^{(i,j)}
    \bm{n}_{ij} + \gamma_{\mathrm{n}} \bm{U}_{\mathrm{n}}^{(i,j)},  \\
    \bm{F}_{\mathrm{C,tan}}^{(i,j)} & = k_{\mathrm{t}} \bm{\xi}^{(i,j)}
    + \gamma_{\mathrm{t}}  \bm{U}_{\mathrm{t}}^{(i,j)} , \\
    \bm{T}_{\mathrm{C,tan}}^{(i,j)} & = a_i \bm{n}_{ij} \times
    \bm{F}_{\mathrm{C,tan}}^{(i,j)},
  \end{split}
\end{equation}
These forces must fulfill Coulomb's law of friction:
$|\bm{F}_{\mathrm{C,tan}}^{(i,j)}| \leq \mu |\bm{F}_{\mathrm{C,nor}}^{(i,j)}|$. 
In the above expression,
$k_{\mathrm{n}}$ and $k_{\mathrm{t}}$ are respectively 
the normal and tangential spring constants, 
and $\gamma_{\mathrm{n}}$ and $\gamma_{\mathrm{t}}$ are the damping constants. 
The normal and tangential relative velocities between two particles $i$ and $j$
are
$\bm{U}_{\mathrm{n}}^{(i,j)} \equiv \operatorPn \bigl(\bm{U}^{(j)} - \bm{U}^{(i)}\bigr)$ 
and 
$\bm{U}_{\mathrm{t}}^{(i,j)} \equiv \operatorPt \Bigl[\bm{U}^{(j)} - \bm{U}^{(i)} 
  - \bigl( a_i \bm{\Omega}^{(i)} + a_j \bm{\Omega}^{(j)} \bigr)\times \bm{n}_{ij}\Bigr] $.
Finally, the quantity $\bm{\xi}^{(i,j)}$ is the tangential spring stretch. 


The computation of the tangential spring stretch $\bm{\xi}^{(i,j)}$,
described in the following, requires some care, 
as we have to impose Coulomb's law, 
which is made difficult by the overdamped dynamics.
At the time $t_{0}$ at which the contact $(i,j)$ is created, 
we set an unstretched tangential spring $ \bm{\xi}^{(i,j)}(t_{0}) = \bm{0} $.
At any further time step $t$ in the simulation, 
the tangential stretch $\bm{\xi}^{(i,j)}(t)$
is incremented according to the value of a ``test'' force,
$ \bm{F}_{\mathrm{C,tan}}^{\prime(i,j)}(t+\mathrm{d}t) 
= k_{\mathrm{t}}\bm{\xi}^{\prime(i,j)}(t+\mathrm{d}t)
+ \gamma_{\mathrm{t}} \bm{U}_{\mathrm{t}}^{(i,j)}(t+dt) $
with the tentative update of stretch
$ \bm{\xi}^{\prime(i,j)}(t+\mathrm{d}t)
= \bm{\xi}^{(i,j)}(t) + \bm{U}_{\mathrm{t}}^{(i,j)}(t) \mathrm{d}t$.
If $\bigl|\bm{F}_{\mathrm{C,tan}}^{\prime(i,j)}(t+\mathrm{d}t)\bigr| 
\leq \mu \bigl|\bm{F}_{\mathrm{C,nor}}^{(i,j)}(t+\mathrm{d}t)\bigr|$, 
the contact is in a static friction state 
and we update the spring stretch as 
$ \bm{\xi}^{(i,j)}(t+\mathrm{d}t) = \bm{\xi}^{\prime(i,j)}(t+\mathrm{d}t)$ 
and the corresponding tangential force is 
$ \bm{F}_{\mathrm{C,tan}}^{(i,j)}(t+\mathrm{d}t) =
\bm{F}_{\mathrm{C,tan}}^{\prime(i,j)}(t+\mathrm{d}t) $.
However, if 
$ \bigl|\bm{F}_{\mathrm{C,tan}}^{\prime(i,j)}(t+\mathrm{d}t)\bigr| 
> \mu \bigl|\bm{F}_{\mathrm{C,nor}}^{(i,j)}(t+\mathrm{d}t)\bigr| $, 
the contact is in sliding state and the spring is updated as 
$ \bm{\xi}^{(i,j)}(t+\mathrm{d}t) = 
(1/k_{\mathrm{t}})\Bigl( \mu \bigl|\bm{F}_{\mathrm{C,nor}}^{(i,j)}(t+\mathrm{d} t)\bigr|
\tanVec - \gamma_{\mathrm{t}} \bm{U}_{\mathrm{t}}^{(i,j)}(t)\Bigr) $,
where the direction is the same as the test force, i.e.,
$\tanVec \equiv \bm{F}_{\mathrm{C,tan}}^{\prime(i,j)} (t+\mathrm{d} t)
/ \bigl|\bm{F}_{\mathrm{C,tan}}^{\prime(i,j)}(t+\mathrm{d} t)\bigr| $.
\emph{Assuming that the velocities are continuous in time},
this ensures that the contact force will at most weakly violate  
the Coulomb friction law for the next time step,
as the force effectively used in the equation of motion will be 
$ \bm{F}_{\mathrm{C,tan}}^{(i,j)}(t+\mathrm{d}t) = 
k_{\mathrm{t}} \bm{\xi}^{(i,j)}(t+\mathrm{d}t) + \gamma_{\mathrm{t}}
\bm{U}_{\mathrm{t}}^{(i,j)}(t+\mathrm{d}t) = 
\mu \bigl|\bm{F}_{\mathrm{C,nor}}^{(i,j)}(t+\mathrm{d} t)\bigr|\tanVec 
+ \gamma_{\mathrm{t}} \dot{\bm{U}}_{\mathrm{t}}^{(i,j)} \mathrm{d}t $,
so that
$\bigl|\bm{F}_{\mathrm{C,tan}}^{(i,j)}(t+\mathrm{d}t)\bigr|
 =  \mu \bigl|\bm{F}_{\mathrm{C,nor}}^{(i,j)}(t+\mathrm{d} t)\bigr| 
+ \mathcal{O}(\mathrm{d} t)$. 


Of course, if velocities happen to be discontinuous, 
the Coulomb's law might be violated significantly, 
and this is the source of one problem when merging this contact model 
with the hydrodynamic interaction model.
Indeed, when contacts are created and destroyed during the flow 
(which happens very frequently), 
the overall resistance matrix changes discontinuously, 
as dashpots are switched on and off.
But, as other position-dependent forces are continuous in time, 
solving the force balance equation will lead to a discontinuity in
velocities when a contact forms or breaks.
This, in turn, leads to large violations of Coulomb's law,
and hence to numerical instabilities where contacts keep switching between
static and dynamic cases at every time step.


Another problem occurs when a contact forms 
(i.e., $\bm{\xi}^{(i,j)}(t) = 0$):
as by definition the normal load on the new contact 
$\bigl|\bm{F}_{\mathrm{C,nor}}^{(i,j)}(t+\mathrm{d}t)\bigr|$
vanishes 
(or is very small in the simulation owing to the finite time-step),
the finite test force
$ \bm{F}_{\mathrm{C,tan}}^{\prime(i,j)} (t+\mathrm{d}t) 
\simeq \gamma_{\mathrm{t}} \bm{U}_{\mathrm{t}}^{(i,j)}(t) $
will lead to an immediate rescaling of the tangential spring as 
$ \bm{\xi}^{(i,j)}(t+\mathrm{d}t) \simeq 
- (\gamma_{\mathrm{t}}/k_{\mathrm{t}}) \bm{U}_{\mathrm{t}}^{(i,j)}(t)$;
i.e., an entirely new finite force appears right at contact time.


These two problems actually also exist in simulations of dry granular matter~%
\citep{Walton_1993a},
but they are more acute with an overdamped dynamics 
because of the direct relation between forces and velocities.
In this case, any discontinuity in forces or velocities
rapidly causes numerical instabilities.
A solution is to use a continuously varying damping
$\gamma_{\mathrm{t}}(h_{ij})$ such that
$\gamma_{\mathrm{t}}(h_{ij}=0)=0$, which ensures the continuity of
forces and velocities~\citep{Walton_1993}
at the cost of increased complexity of the model.
The solution that we prefer using here is to eliminate of the tangential
dashpot by setting $\gamma_{\mathrm{t}} = 0$.
In any case, this dashpot has no physical significance, and it is
only used in granular matter simulations as an efficient numerical stabilizer. 
In our already overdamped context, as long as the hydrodynamic resistance
associated with tangential motion is not too small, 
we do not need such an extra resistive stabilizer,
and we do not face any major difficulty by simply dropping this dashpot.
%


We can quantify this assertion by looking at the relaxation times associated with a contact.
Both normal and tangential contacts have relaxation times.
For the normal part, it is the one of a spring and dashpot system,
$ \tau_{\mathrm{n}} = \gamma_{\mathrm{n}}/k_{\mathrm{n}} $. 
For the tangential part, the damping is provided by the hydrodynamic
resistance, which we here simply denote
$\gamma^{\mathrm{H}}_{\mathrm{t}}$, 
so that $\tau_{\mathrm{t}} = \gamma^{\mathrm{H}}_{\mathrm{t}}/k_{\mathrm{t}} $. 
On the one hand, in order to have a stable numerical scheme, 
one should chose these relaxation times large enough compared 
to the time step $\tau_{\mathrm{n}}$, $\tau_{\mathrm{t}} \gg \mathrm{d}t$.
On the other hand, contacts of hard spheres should react instantaneously
to any external load, 
so physics imposes relaxation times smaller than other typical times 
$\tau_{\mathrm{n}}$, $\tau_{\mathrm{t}} \ll 1/\dot{\gamma}$.
For the normal part of the contact, 
we are free to choose the damping $\gamma_{\mathrm{n}}$ to achieve this.
For the tangential part however, 
we check that those scale separations are verified.

\end{document}